\documentclass[11pt,authoryear]{article}
\usepackage[paperheight=11.7in,paperwidth=8.3in]{geometry}
\usepackage{enumerate}
\usepackage[english]{babel}
\usepackage{amssymb}
\usepackage{graphicx}
\usepackage{caption}
\usepackage{subcaption}
\usepackage{amsfonts}
\usepackage{color}
\usepackage{epstopdf}
\usepackage{dsfont}
\usepackage{bm}
\usepackage{appendix}
\usepackage{authblk}
\usepackage[authoryear]{natbib}

\usepackage{amsthm,amsmath,enumitem}
\usepackage{algorithm}
\usepackage[noend]{algpseudocode}
\usepackage[colorlinks,citecolor=blue,urlcolor=blue]{hyperref}

\setcitestyle{round}

\makeatletter
\def\BState{\State\hskip-\ALG@thistlm}
\makeatother

\newcommand{\vect}[1]{\boldsymbol{\mathbf{#1}}}
\newcommand{\argmin}{\operatornamewithlimits{argmin}}
\newcommand{\argmax}{\operatornamewithlimits{argmax}}
\newcommand{\R}{\mathbb{R}}
\newcommand{\M}{\mathcal{M}}
\newcommand{\prm}{\varphi}

\makeatletter
\def\step{%
   \@ifnextchar[ \@myitem{\@noitemargtrue\@myitem[\@itemlabel]}}
\def\@myitem[#1]{\item[#1]\mbox{}}
\makeatother

\title{Smooth Principal Component Analysis over two-dimensional manifolds with an application to Neuroimaging %\thanksref{T1}
}

\author[1,3]{Eardi Lila\thanks{e.lila@maths.cam.ac.uk}}
\author[2]{John A. D. Aston\thanks{j.aston@statslab.cam.ac.uk}}
\author[3]{Laura M. Sangalli\thanks{laura.sangalli@polimi.it}}
\affil[1]{Cambridge Centre for Analysis, University of Cambridge}
\affil[2]{Statistical Laboratory, DPMMS, University of Cambridge}
\affil[3]{MOX, Dipartimento di Matematica, Politecnico di Milano}
\date{}

\begin{document}

\maketitle

\begin{abstract}%
Motivated by the analysis of high-dimensional neuroimaging signals located over the cortical surface, we introduce a novel Principal Component Analysis technique that can handle functional data located over a two-dimensional manifold. For this purpose a regularization approach is adopted, introducing a smoothing penalty coherent with the geodesic distance over the manifold. The model introduced can be applied to any manifold topology, can naturally handle missing data and functional samples evaluated in different grids of points. We approach the discretization task by means of finite element analysis and propose an efficient iterative algorithm for its resolution. We compare the performances of the proposed algorithm with other approaches classically adopted in literature. We finally apply the proposed method to resting state functional magnetic resonance imaging data from the Human Connectome Project, where the method shows substantial differential
variations between brain regions that were not apparent with other approaches.
\end{abstract}

%\begin{keyword}
%\kwd{Functional Data Analysis}
%\kwd{Principal Component Analysis}
%\kwd{Differential regularization}
%\kwd{Functional Magnetic Resonance Imaging}
%\end{keyword}

%\end{frontmatter}

\section{Introduction}\label{sec:intro}

The recent growth of data arising from neuroimaging has led to profound changes in the understanding of the brain. Neuroimaging is a multidisciplinary activity and the role of statistics in its success should not be underestimated. Much of the work to date has been to determine how to use statistical models in high-dimensional settings that arise out of such imaging modalities as functional Magnetic Resonance Imaging (fMRI) and Electroencephalography (EEG). However, it is becoming increasingly clear that there is now a need to incorporate more and more complex information about brain structure and function into the statistical analysis to enhance our present understanding of the brain.

Considerable amounts of the brain signal captured, for example, by fMRI arise from the cerebral cortex. The cerebral cortex is the highly convoluted thin sheet where most neural activity is focused. It is natural to represent this thin sheet as a 2D surface embedded in a 3D space, structured with a 2D geodesic distance, rather than the 3D Euclidean distance within the volume. In fact, functionally distinct areas may be close to each other if measured with Euclidean distance, but due to the highly convoluted morphology of the cerebral cortex, their 2D geodesic distance along the cortical surface can be far greater. While early approaches to the analysis of hemodynamic signals ignore the morphology of the cortical surface, it has now been well established [\cite{glasser2013} and references therein] that it is beneficial to analyze neuroimaging data through the processing of the signals on the cortical surface using surface-constrained techniques. Classical tools such as non-parametric smoothing models have already been adapted to deal with this kind of data, see e.g. \cite{chung2014}.

The goal of the present paper is to introduce a novel Principal Component Analysis (PCA) technique suitable for working with functional signals distributed over curved domains and specifically over two-dimensional smooth Riemannian manifolds, such as the cortical surface. The cortical surface can be extracted from structural Magnetic Resonance Imaging (MRI), a non-invasive scanning technique used to visualize the internal structure of the brain, rendering it as a 3D image with high spatial resolution. The signal of interest, which we want to analyse with respect to the surface, comes from fMRI, which detects a Blood Oxygen Level Dependent (BOLD) signal [\cite{Ogawa90}] as a series of repeated measurements in time, yielding a time series of 3D images. An increased neural activity in a particular area of the brain causes an increased demand for oxygen. As the fMRI signal is related to changes in the relative ratio of oxy- to deoxy-hemoglobin, due to their differing magnetic properties, the signal captured within an fMRI scan is considered to be a surrogate for neural activity and is used to produce activation maps or investigate brain functional connectivity. The fMRI signal of each individual related to the neural activity in the cerebral cortex is generally mapped on a common template cortical surface, to allow multi-subject statistical analysis.

In this paper, in particular, we will focus our attention on functional connectivity (FC). FC maps, on the cortical surface, can be constructed computing the pairwise correlation between all vertex's fMRI time-series and the mean time-series of a region of interest. %The fMRI signal is acquired in the absence of any task and for this reason the associated FC maps are also called resting state functional connectivity maps. %
The resulting FC map for each subject provides a clear view of areas to which the region of interest is functionally connected.

In practice, the template cortical surface is represented by a triangulated surface that can be considered a discrete approximation of the underlying smooth compact two-dimensional Riemannian manifold $\mathcal{M} \subset \mathbb{R}^3$ modelling the cortical surface. Each resting state FC map can be represented by a function $x_i: \mathcal{M} \rightarrow \mathbb{R}$.
Once we have the correlation maps on the cortical surface we want to study how the phenomena varies from subject to subject. A statistical technique for this study is PCA.  It is natural to contextualize this task in the framework of Functional Data Analysis [\cite{ramsay2005}].

In Section~\ref{sec:FPCA} we establish the formal theoretical properties of Functional PCA (FPCA) in the case of random functions whose domain is a manifold $\mathcal{M}$. In Section~\ref{sec:SM-FPCA} we introduce a novel FPCA model and propose an algorithm for its resolution. We then give some simulation results in Section~\ref{sec:simulations}, indicating the performance of our methodology, as compared to other methods in literature. We then return to the FC maps example in Section~\ref{sec:app}, to consider how the surface based PCA analysis might be used in this case and draw some concluding remarks in Section~\ref{sec:discussion}.

\section{Functional principal component analysis}\label{sec:FPCA}
Consider the space of square integrable functions on $\mathcal{M}$: $L^2(\mathcal{M}) = \{f : \mathcal{M} \rightarrow \mathbb{R}: \int_{\mathcal{M}} |f(p)|^2 dp < \infty\}$ with the inner product ${\langle f,g \rangle}_\mathcal{M} = \int_{\mathcal{M}} f(p)g(p) dp$ and norm $\|f\|_{\mathcal{M}}=\int_{\mathcal{M}} |f(p)|^2 dp$. Consider the random variable $X$ with values in $L^2(\mathcal{M})$, mean $\mu = \mathbb{E}[X]$ and a finite second moment, i.e. $\int_{\mathcal{M}} \mathbb{E}[X^2] < \infty $, and assume that its covariance function $K(p, q) = \mathbb{E}[(X(p)-\mu(p))(X(q)-\mu(q))]$ is square integrable. Mercer's Lemma [\cite{riesz1955}] guarantees the existence of a non-increasing sequence $(\kappa_j)$ of eigenvalues of $K$ and an orthonormal sequence of corresponding eigenfunctions $(\psi_j)$, such that
\begin{equation}
\int_{\mathcal{M}} K(p,q) \psi_j(p) dp = \kappa_j \psi_j(q), \qquad \forall q \in \mathcal{M}
\end{equation}
and that $K(p,q)$ can be written as $K(p,q) = \sum_{j=1}^\infty \kappa_j\psi_j(p)\psi_j(q)$ for each $p,q \in \mathcal{M}$. Thus $X$ can be expanded as $X = \mu + \sum_{j=1}^\infty \varepsilon_j \psi_j$, where the random variables $\varepsilon_1,\varepsilon_2,\ldots$ are uncorrelated and are given by $\varepsilon_j = \int_{\mathcal{M}} \{X(p) - \mu(p) \}\psi_j(p)dp$. This is also known as the Karhunen-Lo\`{e}ve (KL) expansion of $X$.

The collection $(\psi_j)$ defines the strongest modes of variation in the random function $X$ and these are called Principal Component (PC) functions. In fact $\psi_1$ is such that
\begin{equation*}
\psi_1 = \argmax_{\phi:\|\phi\|_{\mathcal{M}}=1}\int_{\mathcal{M}}\int_{\mathcal{M}}\phi(p)K(p,q)
\phi(q)dpdq,
\end{equation*}
while $\psi_m$, for $m>1$, solves an analogous problem with the added constraint of $\psi_m$ being orthogonal to the previous $m-1$ functions $\psi_1, \ldots, \psi_{m-1}$, i.e.
\begin{equation*}
\psi_m = \argmax_{\scriptsize \begin{array}{clc}\phi:\|\phi\|_{\mathcal{M}}=1 \\ \langle \phi, \psi_j \rangle_{\mathcal{M}} = 0 \quad j = 1,\ldots,m-1 \end{array}} \int_{\mathcal{M}}\int_{\mathcal{M}}\phi(p)K(p,q)
\phi(q)dpdq.
\end{equation*}
The random variables $\varepsilon_1,\varepsilon_2,\ldots$ are called PC scores.

Another important property of PC functions is the best $M$ basis approximation. In fact, for any fixed $M \in \mathbb{N}$, the first $M$ PC functions of $X$ satisfies
\begin{equation}\label{eq:minimization}
{(\psi_i)}_{m=1}^M =
\argmin_{(\{\phi_m\}_{m=1}^M: \langle \phi_m, \phi_l \rangle = \delta_{ml})}
\mathbb{E}\int_{\mathcal{M}}\bigg\{X-\mu - \sum_{m=1}^M \langle X, \phi_m \rangle \phi_m \bigg\}^2,
\end{equation}
where $\delta_{ml}$ is the Kronecker delta; i.e. $\delta_{ml}=1$ for $m=l$ and $0$ otherwise.

Suppose $x_1, \ldots , x_n$ are $n$ smooth samples from $X$. Usually, for each of these functions, only noisy evaluations $x_i(p_j)$ on a fixed discrete grid of points $p_1,\ldots,p_s$ are given. In this setting, we will now recall the two standard approaches to FPCA: the pre-smoothing approach and the regularized PCA approach.

The pre-smoothing approach is based on the two following steps. In the first step, the observations associated to each function are smoothed, in order to obtain smooth representations of $x_1, \ldots , x_n$. Then, the sample mean $\bar{x} = n^{-1} \sum_i x_i$ and the sample covariance $\hat{K}(p, q) = \frac{1}{n} \sum_{i=1}^n(x_i(p)-\bar{x}(p))(x_i(q)-\bar{x}(q))$ are used to estimate $\mu$ and $K$ respectively. % Applying the orthonormal basis expansion to $\hat{K}$ we can write $\hat{K}(p,q) = \sum_{j=1}^\infty \hat{\kappa}_j \hat{\psi}_j(p)\hat{\psi}_j(q)$ with $p,q \in \mathcal{M}$.
%The sequence $\hat{\psi_1},\hat{\psi_2},\ldots$ is usually treated as an approximation of $\psi_1,\psi_2,\ldots$.
Finally, the estimates of the PC functions $\hat{\psi_1},\hat{\psi_2},\ldots$ are computed through the characterization $\int_{\mathcal{M}} \hat{K}(p,q) \hat{\psi}_j(p) dp = \hat{\kappa}_j \hat{\psi}_j(q)$, which is solved by the discretization of the problem on a fine grid or by the basis expansion of estimated smooth functions. In the case where the domain is an interval of the real line, a theoretical study on the accuracy of $\hat{\psi_j}$ as an estimate of $\psi_j$ is offered for example in \cite{hall2006}.

%Consider now the empirical version of (\ref{eq:minimization})
%\begin{equation}\label{eq:best_basis}
%{(\psi_i)}_{i=1}^M =
%\argmin_{(\{\phi_m\}_{m=1}^M: \langle \phi_m, \phi_l \rangle_\mathcal{M} = \delta_{ml})}
%\frac{1}{n} \sum_{i=1}^n \int_{\mathcal{M}}\bigg\{x_i-\mu - \sum_{m=1}^M \langle x_i, \phi_m \rangle \phi_m \bigg\}^2.
%\end{equation}

Define the $n \times s$ matrix $\vect{X} = (x_i(p_j))$, the column vector $\vect{\mu} = (\frac{1}{n}\sum_{i=1}^n x_i(p_j))$ of length $s$, the $n \times M$ matrix $\vect{A} = (\langle X_i, \phi_m \rangle)$ and the $s \times M$ matrix $\vect{\Phi} = (\phi_m(p_j))$. Let $\vect{1}$ denote the column vector of length $n$ with all entries equal to $1$. The empirical counterpart of the objective function in (\ref{eq:minimization}) becomes
\begin{equation}\label{eq:SVD}
\frac{1}{n} \| \vect{X} - \vect{1} \vect{\mu}^T - \vect{A} \vect{\Phi}^T \|_F^2,
\end{equation}
where $\| \cdot \|_F$ is the Frobenius norm, defined as the square root of the sum of the squares of its elements.
This last formulation gives a natural way to deal with the fact that only pointwise and noisy evaluations $x_i(p_j), \, i=1,\ldots,n, j=1,\ldots,s$ of the underlying functional samples are usually available. However, it does not incorporate any information on the smoothness of the functional data. In fact, considering the Singular Value Decomposition (SVD) of $\vect{X} - \vect{1} \vect{\mu}^T = \vect{U} \vect{D} \vect{V}^T$, it can be shown that the minimizing arguments of (\ref{eq:SVD}) are $\vect{\hat{\Phi}} = \vect{V}$ and $\vect{\hat{A}} = \vect{U} \vect{D}$, thus the obtained formulation is a multivariate PCA applied to the data-matrix $\vect{X}$.

The regularized PCA approach consists on adding a penalization term to the classic formulation of the PCA, in order to recover a desired feature of the estimated underlying functions. In particular the formulation (\ref{eq:SVD}) has shown a great flexibility for this purpose. Examples of models where a sparseness property is assumed on the data are offered for instance in \cite{jolliffe2003,zou2005,shen2008}. In the specific case of functional data analysis, the penalization term usually encourages the PC functions to be smooth.  Examples of PCA models that explicitly incorporates a smoothing penalization term are given by \cite{rice1991,silverman1996,huang2008}. The cited works deal with functions whose domain is a limited interval in $\mathbb{R}$, and in particular, our proposal can be seen as an extension of \citet{huang2008} to the case of functions whose domain is a two-dimensional manifold. \cite{zhou2014} recently proposed a smooth FPCA for two-dimensional functions on irregular planar domains; their approach is based on a mixed effects model that specifies the PC functions as bivariate splines on triangulations and the PC scores as random effects. Here we propose a FPCA model that can handle real functions observable on a two-dimensional manifold. We shall consider a smoothing penalty operator, coherent with the 2D geodesic distances on the manifold. This leads to the definition of a model that can fully exploit the information about the geometry of the manifold.

\section{Smooth FPCA over two-dimensional manifolds}\label{sec:SM-FPCA}
\subsection{Geometric concepts}
\begin{figure}[h]
\centering
\includegraphics[width=.9\textwidth]{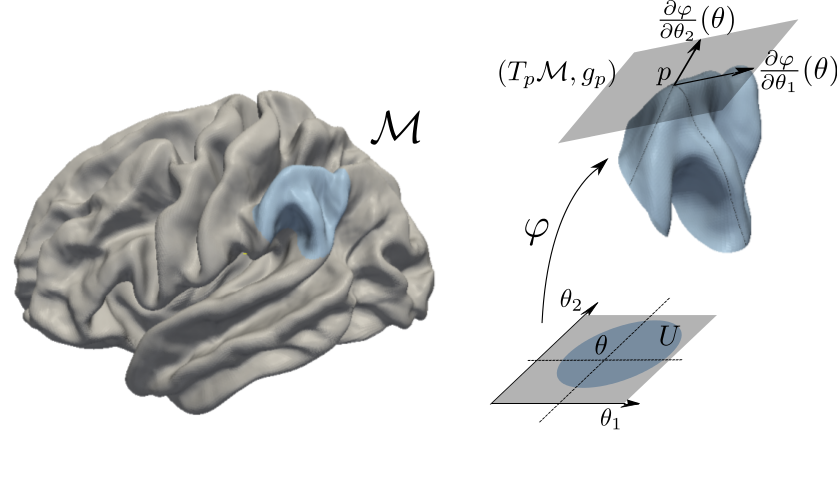}
\caption{A pictorial representation of the geometric objects modelling the idealized cortical surface $\mathcal{M}$.}
\label{fig:geometric_concepts}
\end{figure}

We first introduce the essential geometric concepts that allow the definition of the Laplace-Beltrami operator, which plays a central role in the proposed model. In detail, let the bijective and smooth function $\prm: U \subset \R^2 \rightarrow \R^3$ be a local parametrization of $\M$ around the point $p \in \M$, as depicted in Figure~\ref{fig:geometric_concepts}. Let $\theta \in U$ be such that $\theta = \prm^{-1}(p) $, then %$\big \{ \frac{\partial \prm}{\partial \theta_i}(\theta) \}_{i=1,2}$
\begin{equation}\label{eq:basis_tangent}
\big \{\frac{\partial \prm}{\partial \theta_i}(\theta) \}_{i=1,2}
\end{equation}
defines a basis for the tangent space $T_p \M$ at the point $p$.

The Riemannian manifold $\M$ can be equipped with a metric by defining a scalar product $g_p$ on the tangent space $T_p \M$. This enables, for instance, the computation of the lengths of curves or integrals on the surface. Fixing the reference system on the tangent plane with the basis (\ref{eq:basis_tangent}), we can represent $g_p$ as the matrix $G = (g_{ij})_{i,j=1,2}$ such that
\begin{equation*}
g_p(v,w) = \sum_{i,j=1}^2 g_{ij} v_i w_j
\end{equation*}
for all $v = \sum v_i \frac{\partial \prm}{\partial \theta_i}(\theta)$ and $w = \sum w_i \frac{\partial \prm}{\partial \theta_i}(\theta)$. In our case it is natural to consider the scalar product induced by the Euclidean embedding space $\R^3$, i.e. the first fundamental form
\begin{equation*}
g_{ij}(\theta) = \frac{\partial \prm}{\partial \theta_i}(\theta) \cdot \frac{\partial \prm}{\partial \theta_j}(\theta),
\end{equation*}
where $\cdot$ denotes the inner product in $\R^3$. Moreover, we denote by $G^{-1} = (g^{ij})_{i,j=1,2}$ the inverse of the matrix $G$ and by $g = det(G)$ the determinant of the matrix $G$.

Let now $f: \M \rightarrow \R$ be a real valued and twice differentiable function on the manifold $\M$.
% and let $df_p(v)$ be its directional derivative at $p$ in the direction $v$. The concept of directional derivative can be used to define $(\nabla_\M f)(p) \in T_p\M$, the gradient of the function $f$ at $p$, as the element of the tangent space that satisfies $g_p(\nabla_\M f(p),v) = df_p(v)$ for all $v \in T_p\M$, which is a well-known property of the gradient in Euclidean spaces. Let $F = f \circ \prm$, then it can be shown that $(\nabla_\M f)(p)$ can be expressed in local coordinates as
Let $F = f \circ \prm$, then the gradient $\nabla_\M f$ is defined as
\[
(\nabla_\M f)(p) = \sum_{i,j=1}^2 g^{ij}(\theta) \frac{\partial F}{\partial \theta_j}(\theta) \frac{\partial \prm}{\partial \theta_j}(\theta).
\]
In the case of a flat manifold $\M$, the last expression reduces to the expression of the gradient in $\R^2$, i.e. $\nabla = (\frac{\partial}{\partial \theta_1},\frac{\partial}{\partial \theta_2})$

%Let $V$ be a smooth vector field on the closed manifold $\M$, where $V(p) \in T_p\mathcal{M}$. A generalization of the divergence operator $\div_\M$ can be introduced by imposing that $ \int_\M f \div_\M V = - \int_\M g_p (\nabla_\M f, V)$  for all $f \in C^\infty(\M)$, this also a  well-known property in the Euclidean case. As in the Euclidean case, we can finally define the Laplace-Beltrami operator as $\Delta_\M f = \div_\M \nabla_\M f$. In local coordinates it can be shown that $\Delta_\M f(p)$ is of the form
The Laplace-Beltrami operator $\Delta_\M$ is a generalization to the case of surfaces of the standard Laplacian defined on $\R^n$, i.e. $\Delta = \sum_{i=1}^n \frac{\partial^2}{\partial^2 \theta_i}$. It is related to the second partial derivatives of $f$ on $\mathcal{M}$, i.e. its local curvature, and it is defined as
\begin{equation*}
(\Delta_\M f)(p) = \frac{1}{\sqrt{g(\theta)}} \sum_{i,j=1}^2 \frac{\partial}{\partial \theta_j} g^{ij} \sqrt{g(\theta)} \frac{\partial F}{\partial \theta_j}(\theta).
\end{equation*}
The defined operator is invariant with respect to rigid transformations of the reference system on $U$.

\subsection{Model}
Suppose now the sample of $n$ functions $x_i: \mathcal{M} \rightarrow \mathbb{R}$ is observed at a fixed set of points ${p_1, \ldots, p_s}$ in $\mathcal{M}$ (this will be relaxed later). %As before, let $f:\M \rightarrow \R$ be a real valued and twice differentiable function and
Let $\vect{u} = \{u_i\}_{i=1,\ldots,n}$ be a n-dimensional real column vector. We propose to estimate the first PC function $\hat{f}:\mathcal{M}\rightarrow\mathbb{R}$ and the associated PC scores vector $\hat{\vect{u}}$, by solving the equation
\begin{equation}\label{eq:model}
(\hat{\vect{u}},\hat{f}) = \argmin \limits_{\vect{u},f} \sum \limits_{i=1}^n \sum \limits_{j=1}^s (x_i(p_j) - u_i f(p_j))^2 + \lambda \vect{u}^T\vect{u} \int_{\mathcal{M}} \! \Delta^2_{\mathcal{M}} f,
\end{equation}
where the Laplace-Beltrami operator is integrated over the manifold $\mathcal{M}$, enabling a global roughness penalty on $f$, while the empirical term encourages $f$ to capture the strongest mode of variation. The parameter $\lambda$ controls the trade-off between the empirical term of the objective function and roughness penalizing term. The $\vect{u}^T\vect{u}$ term is justified by some invariance considerations on the objective function as done in the case of one dimensional domains, in \cite{huang2008}. % As before, let $\vect{X}$ denote the $n \times s$ matrix $\vect{X}=(x_i(p_j))$ and %
Consider the transformation $(\vect{u} \rightarrow c\vect{u}, f \rightarrow \frac{1}{c}f)$, with $c$ a constant, and the transformation $(\vect{X} \rightarrow c \vect{X}, \vect{u} \rightarrow c\vect{u})$, where $\vect{X} = (x_i(p_j))$. Then the objective function in (\ref{eq:model}) is invariant with respect to the first transformation, while the empirical and the smoothness terms are re-scaled by the same coefficient with the second transformation.

The subsequent PCs can be extracted sequentially by removing the preceding estimated components from the data matrix $\vect{X}$. This allows the selection of a different penalization parameter $\lambda$ for each PC estimate. We will refer to the model introduced as Smooth Manifold FPCA (SM-FPCA).

\subsection{Iterative algorithm}

Here we present the numerical algorithm for the resolution of the model introduced above. Our approach for the minimization of the functional (\ref{eq:model}) can be summarized in the following two steps:
\begin{itemize}
\item Splitting the optimization in a finite dimensional optimization in $\vect{u}$ and an infinite-dimensional optimization in $f$;
\item Approximating the infinite-dimensional solution thanks to a Surface Finite Element discretization.
\end{itemize}

Let $\vect{f}_s$ be the vector of length $s$ such that $\vect{f}_s = (f(p_1), \ldots, f(p_s))^T$. The expression in (\ref{eq:model}) can be rewritten as
\begin{equation}\label{eq:model_matricial}
(\hat{\vect{u}},\hat{f}) = \argmin \limits_{\vect{u},f} \|\vect{X} - \vect{u} \vect{f}^T_s \|_F^2 + \lambda \vect{u}^T\vect{u} \int_{\mathcal{M}} \! \Delta^2_{\mathcal{M}} f.
\end{equation}
A normalization constraint must be considered in this minimization problem to make the representation unique, as in fact multiplying $\vect{u}$ by a constant and dividing $f$ by the same constant does not change the objective function (\ref{eq:model_matricial}). In particular we set the constraint $\|\vect{u}\|_2 = 1$, as this allows us to leave the infinite-dimensional optimization in $f$ unconstrained.

Our proposal for the minimization of the criterion (\ref{eq:model_matricial}) is to alternate the minimization of $\vect{u}$ and $f$ in an iterative algorithm:

\begin{enumerate}[label=\textit{Step} \arabic*,align=left, leftmargin=1.0cm]

\step Estimation of $\vect{u}$ given $f$. For a given $f$, the minimizing $\vect{u}$ of the objective function in (\ref{eq:model_matricial}) is
\begin{equation}\label{eq:u_full}
\vect{u} = \frac{\vect{X}\vect{f}_s}{\| \vect{f}_s \|_2^2 +  \lambda \int_{\mathcal{M}} \! \Delta^2_{\mathcal{M}} f},
\end{equation}
and the minimizing unitary-norm vector $\vect{u}$ is
\begin{equation}\label{eq:u}
\vect{u} = \frac{\vect{X}\vect{f}_s}{\|\vect{X}\vect{f}_s\|_2}.
\end{equation}

\step Estimation of $f$ given $\vect{u}$. For a given $\vect{u}$, solving (\ref{eq:model_matricial}) with respect to $f$ is equivalent to finding the $f$ that minimizes

\begin{equation}\label{eq:f_min}
J_{\lambda, \vect{u}}(f) = \vect{f}_s^T\vect{f}_s+\lambda \int_{\mathcal{M}} \! \Delta^2_{\mathcal{M}} f -2\vect{f}_s^T\vect{X}^T\vect{u}.
\end{equation}

\end{enumerate}

%% Piano: Ramsey(2002), Wood(2008), Bivariate Spline, Spherical: Wahba (1981), Alfeld et al (1996), Baramidze (2006)
Step 1 is basically the classical expression of the score vector given the loadings vector, where in this case the loading vector is given by $\vect{f}_s$, the evaluations of the PC function in $p_1,\ldots,p_s$. The problem in Step 2 is not trivial, consisting in an infinite-dimensional minimization problem. Let $z_j$ denote the $j$th element of the vector $\vect{X}^T \vect{u}$, then minimizing the functional in (\ref{eq:f_min}) is equivalent to minimizing
\begin{equation}\label{eq:f_reg}
\sum_{j=1}^{s} \bigg(z_j - f(p_j)\bigg)^2 +\lambda \int_{\mathcal{M}} \! \Delta^2_{\mathcal{M}} f.
\end{equation}
This problem involves estimating a smooth field $f$ defined on a manifold, starting from noisy observations $z_j$ at points $p_j$. In the case of real functions defined on the real line, adopting a penalty of the form $\lambda \int f''$, the minimization problem turns out to have a finite-dimensional closed form solution that is a cubic spline [\cite{Silv}]. For real functions defined on an Euclidean space, cubic splines are generalized by thin-plate splines. In this case, for an opportune smoothing penalty, the solution of the minimization problem can be expressed in terms of a finite linear combination of radial basis functions [\cite{Duchon1977}].

However, the case of real functions defined on a non-Euclidean domain $\M$ is more involved. In the special case where $\mathcal{M}$ is a sphere or a sphere-like surface, that is $\mathcal{M} = \{\sigma(v) = \rho(v)v : v \in S \}$ where $S \subset \mathbb{R}^3$ is the unit sphere centered at the origin, this smoothing problem has been considered, among others, by \cite{wahba1981} and \cite{alfeld1996}. Moreover, the functional (\ref{eq:f_reg}) is considered, among others, by \cite{SSRM1} and \cite{SSRM2}. Here $\mathcal{M}$ is respectively a manifold homeomorphic to an open ended cylinder and a manifold homeomorphic to a sphere. In the latter two works the field $f$ is estimated by first conformally recasting the problem to a planar domain and then discretizing it by means of planar finite elements, generalizing the planar smoothing model in \cite{ramsay2002}. Our approach is also based on a Finite Element (FE) discretization, but differently from \cite{SSRM1} and \cite{SSRM2}, we construct here a FE space directly on the triangulated surface $\mathcal{M}_{\mathcal{T}}$ that approximates the manifold $\mathcal{M}$, i.e. we use surface FE, avoiding any flattening step and thereby allowing the formulation to be applicable to any manifold topology.

\subsection{Surface Finite Element discretization}
Assume, for clarity of exposition only, that $\mathcal{M}$ is a closed surface, as in our motivating application. The case of non-closed surfaces can be handled by considering some appropriate boundary conditions as done for instance in the planar case in \cite{sangalli2013}. Consider the linear functional space $H^2(\mathcal{M})$, the space of functions in $L^2(\mathcal{M})$ with first and second weak derivatives in $L^2(\mathcal{M})$. The infinite dimensional part of the estimation problem can be reformulated as follows: find $\hat{f} \in H^2(\mathcal{M})$ such that
\begin{equation}\label{eq:problem}
\hat{f}=\argmin_{f \in H^2(\mathcal{M})} J_{\lambda, \vect{u}}(f).
\end{equation}

\textbf{Proposition 1.} \textit{The solution $\hat{f} \in H^2(\mathcal{M})$ exists and is unique and is such that
\begin{equation} \label{eq:eulero-lagrange}
\sum_{j=1}^s \varphi(p_j)\hat{f}(p_j) + \lambda \int_\mathcal{M} \Delta_\mathcal{M} \varphi \Delta_\mathcal{M} \hat{f} = \sum_{j=1}^s \varphi(p_j) \sum_{i=1}^n x_i(p_j)u_i
\end{equation}
for every $\varphi \in H^2(\mathcal{M})$.}

As detailed in the Supplementary Material, the key idea is to minimize $J_{\lambda, \vect{u}} (f)$ by differentiating this functional with respect to $f$. This leads to (\ref{eq:eulero-lagrange}), that characterizes the estimate $\hat{f}$ as the solution of a linear fourth-order problem.

\begin{figure}[h]
\centering
\begin{subfigure}{.5\textwidth}
  \centering
  \includegraphics[width=.9\textwidth]{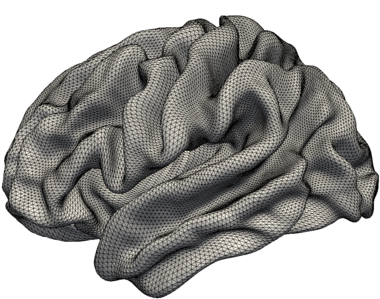}
\end{subfigure}%
\begin{subfigure}{.5\textwidth}
  \centering
  \includegraphics[width=.9\textwidth]{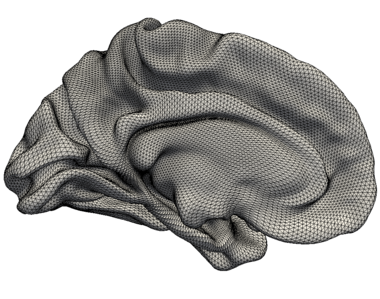}
\end{subfigure}
\caption{The triangulated surface approximating the left hemisphere of the template brain. The mesh is composed by 32K nodes and by 64K triangles}
\label{fig:brain_mesh}
\end{figure}

Consider now a triangulated surface $\mathcal{M}_\mathcal{T}$, union of the finite set of triangles $\mathcal{T}$, giving an approximated representation of the manifold $\mathcal{M}$. Figure~\ref{fig:brain_mesh} for instance shows the triangulated surface approximating the left hemisphere of a template brain. We then consider the linear finite element space $V$ consisting in a set of globally continuous functions over $\mathcal{M}_\mathcal{T}$ that are linear affine where restricted to any triangle $\tau$ in $\mathcal{T}$, i.e.
\begin{equation*}
V = \{ v \in C^0(\mathcal{M}_\mathcal{T}): v|_{\tau} \text{ is linear affine for each } \tau \in \mathcal{T} \}.
\end{equation*}

This space is spanned by the nodal basis $\psi_1, \ldots, \psi_K$  associated to the nodes $\xi_1,\ldots,\xi_K$, corresponding to the vertices of the triangulation $\mathcal{M}_\mathcal{T}$. Such basis functions are lagrangian, meaning that $\psi_i(\xi_j)=1$ if $i=j$ and $\psi_i(\xi_j)=0$ otherwise. Setting $\vect{f} = (f(\xi_1),\ldots,f(\xi_K))^T$ and $\vect{\psi} = (\psi_1,\ldots,\psi_K)^T$, every function $f \in V$ has the form
\begin{equation}\label{eq:basis}
f(p) = \sum_{k=1}^K f(\xi_k) \psi_k(p) = \vect{f}^T \vect{\psi}(p)
\end{equation}
for each $p \in \mathcal{M}_\mathcal{T}$. The surface finite element space provides a finite dimensional subspace of $H^1(\mathcal{M})$ [\cite{dziuk}]. To use this finite element space to discretize the infinite-dimensional problem (\ref{eq:eulero-lagrange}), that is well posed in $H^2(\mathcal{M})$, we first need a reformulation of (\ref{eq:eulero-lagrange}) that involves only first-order derivatives. This can be obtained by introducing an auxiliary function $g$ that plays the role of $\Delta_\mathcal{M} f$, splitting the equation (\ref{eq:eulero-lagrange}) into a coupled system of second-order problems and finally integrating by parts the second order terms. The details of this derivation can be found in the supplementary material.
The discrete estimators $\hat{f}_h, \hat{g}_h \in V$ are then obtained by solving
{\small
\begin{align}\label{eq:discrete}
\begin{cases}
& \int_{\mathcal{M}_\mathcal{T}} \nabla_{\mathcal{M}_\mathcal{T}} \hat{f}_h \nabla_{\mathcal{M}_\mathcal{T}} \varphi_h - \int_{\mathcal{M}_\mathcal{T}} \hat{g}_h \varphi_h = 0\\
& \lambda \!\int_{\mathcal{M}_\mathcal{T}} \nabla_{\mathcal{M}_\mathcal{T}} \hat{g}_h \nabla_{\mathcal{M}_\mathcal{T}} v_h + \sum\limits_{j=1}^s \hat{f}_h(p_j)v_h(p_j) = \sum\limits_{j=1}^s v_h(p_j) \sum\limits_{i=1}^n x_i(p_j)u_i
\end{cases}
\end{align}
}
for all $\varphi_h, v_h \in V$. Define the $s \times K$ matrix $\vect{\Psi} = (\psi_k(p_j))$ and the $K \times K$ matrices $\vect{R}_0=\int_{\mathcal{M}_{\mathcal{T}}}(\vect{\psi} \vect{\psi}^T)$ and $\mathbf{R}_1=\int_{\mathcal{M}_\mathcal{T}} (\nabla_{\mathcal{M}_\mathcal{T}} \vect{\psi}) (\nabla_{\mathcal{M}_\mathcal{T}} \vect{\psi})^T$. Then, exploiting the representation (\ref{eq:basis}) of functions in $V$ we can rewrite (\ref{eq:discrete}) as a linear system. Specifically the Finite Element solution $\hat{f}_h(p)$ of the discrete counterpart (\ref{eq:discrete}) is given by $\hat{f}_h(p) = \vect{\psi}(p)^T \vect{\hat{f}}$ where $\vect{\hat{f}}$ is the solution of
\begin{equation}\label{eq:linear_system}
	\begin{bmatrix}
		\vect{\Psi}^T\vect{\Psi}& \lambda\vect{R}_{1}\\
		\lambda\vect{R}_{1}& -\lambda\vect{R}_{0}
	\end{bmatrix}
	\begin{bmatrix}
		\vect{\hat{f}}\\
		\vect{\hat{g}}
	\end{bmatrix}
=
	\begin{bmatrix}
		\vect{\Psi}^T\vect{X}^T\vect{u}\\
		\vect{0}
	\end{bmatrix}
\end{equation}
Solving (\ref{eq:linear_system}) leads to
\begin{equation}
\vect{\hat{f}} = (\vect{\Psi}^T\vect{\Psi} + \lambda \vect{R}_{1} \vect{R}_{0}^{-1} \vect{R}_{1})^{-1} \vect{\Psi}^T \vect{X}^T\vect{u}.
\end{equation}
Although this last formula is a compact expression of the solution, it is preferable to compute the solution from the linear system (\ref{eq:linear_system}) due to the sparsity property of the matrix in the left-hand side. As an example, in the simulations and the application shown in Sections~\ref{sec:simulations}-\ref{sec:app}, respectively less then $1\%$ and less then $0.1\%$ of the elements in the matrix in the left hand side of (\ref{eq:linear_system}) are different from zero, allowing a very efficient solution of the linear system.

In the model introduced, we assume that all the observed functions $x_i$ are sampled on the common set of points  $p_1, \ldots, p_s \in \mathcal{M}$. Suppose moreover, $p_1, \ldots, p_s \in \mathcal{M}$ coincide with the vertices of the triangulated surface $\mathcal{M}_\mathcal{T}$. In this particular case, an alternative approach could consist of interpreting the points $p_1, \ldots, p_s \in \mathcal{M}_\mathcal{T}$ as the nodes of a graph linked by the edges of the triangulation and considering the model (\ref{eq:model}) with a discrete smoothness operator term instead of the Laplace-Beltrami operator (see e.g. \cite{belkin2001} for the choice of the penalization term and \cite{deng2011} for an application to matrix decomposition). However, thanks to its functional nature, the formulation (\ref{eq:model}) can be easily extended to the case of missing data or sparsely sampled functional data. Specifically, suppose now that each function $x_i$ is observable on a set of points $p_1^{i}, \ldots, p_{s_i}^{i}$, then the natural extension of the model (\ref{eq:model}) becomes
\begin{equation}\label{eq:model_sparsedata}
(\hat{\vect{u}},\hat{f}) = \argmin \limits_{\vect{u},f} \sum \limits_{i=1}^n \sum \limits_{j=1}^{s_i} (x_i(p_j^{i}) - u_i f(p_j^{i}))^2 + \lambda \vect{u}^T\vect{u} \int_{\mathcal{M}} \! \Delta^2_{\mathcal{M}} f.
\end{equation}
Following the same procedure, we can define an analogous algorithm based on the following two steps.
\begin{enumerate}[label=\textit{Step} \arabic*,align=left, leftmargin=1.0cm]

\step For a given $f$, the unitary-norm vector $\vect{u}$ minimizing (\ref{eq:model_sparsedata}) is given by
\begin{equation*}
\vect{u} \text{ such that } u_i = \frac{\
\sum_{j=1}^{s_i} x_i(p_j^{i}) f(p_j^{i})}{\surd \sum_{i=1}^n (\sum_{j=1}^{s_i} x_i(p_j^{i}) f(p_j^{i}))^2}.
\end{equation*}

\step For a given $\vect{u}$, the function $f$ minimizing (\ref{eq:model_sparsedata}) is given by\\
$f = \vect{f}^T \vect{\psi}$ with $\vect{f}$ such that
\begin{equation*}
	\begin{bmatrix}
		\vect{L}& \lambda\vect{R}_{1}\\
		\lambda\vect{R}_{1}& -\lambda\vect{R}_{0}
	\end{bmatrix}
	\begin{bmatrix}
		\vect{f}\\
		\vect{g}
	\end{bmatrix}
=
	\begin{bmatrix}
		\vect{D}^T\vect{u}\\
		\vect{0}
	\end{bmatrix},
\end{equation*}
where
\begin{equation*}
\begin{array}{clc}
\vect{L} &=
\begin{bmatrix}
\sum \limits_{i=1}^n \sum \limits_{j=1}^{s_i} u_i^2 \psi_1(p_j^{i}) \psi_1({p}_j^i)
\quad \ldots \quad
\sum \limits_{i=1}^n \sum \limits_{j=1}^{s_i} u_i^2 \psi_1(p_j^{i}) \psi_K({p}_j^i)\\
\ldots & \\
\sum \limits_{i=1}^n \sum \limits_{j=1}^{s_i} u_i^2 \psi_K(p_j^{i}) \psi_1({p}_j^i)
\quad \ldots \quad
\sum \limits_{i=1}^n \sum \limits_{j=1}^{s_i} u_i^2 \psi_K(p_j^{i}) \psi_K({p}_j^i)
\end{bmatrix}\\
\vect{D} &=
\begin{bmatrix}
\sum \limits_{j=1}^{s_1} \psi_1({p}_j^1) x_1({p}_j^1)
\quad \ldots \quad
\sum \limits_{j=1}^{s_n} \psi_1({p}_j^n) x_n({p}_j^n) \\
\ldots & \\
\sum \limits_{j=1}^{s_1} \psi_K({p}_j^1) x_1({p}_j^1)
\quad \ldots \quad
\sum \limits_{j=1}^{s_n} \psi_K({p}_j^n) x_n({p}_j^n)
\end{bmatrix}.
\end{array}
\end{equation*}
\end{enumerate}

\subsection{SM-FPCA Algorithm}
The algorithm for the resolution of the model SM-FPCA (\ref{eq:model}) can be summarized in the following steps.

\begin{algorithm}[H]
\caption{SM-FPCA Algorithm}\label{alg:fPCA}
\begin{algorithmic}[1]
\State Initialization:
\begin{enumerate}[label=(\alph*)]
\item Computation of $\vect{\Psi}$, $\vect{R}_0$ and $\vect{R}_1$
\item Perform the SVD: $\vect{X} = \vect{UDV}^T$
\item $\vect{f}_s \gets \vect{V}[:,1]$, where $\vect{V}[:,1]$ are the loadings of the first PC
\end{enumerate}
\State Scores estimation:
\begin{equation*}
\vect{u} \gets \frac{\vect{X}\vect{f}_s}{\|\vect{X}\vect{f}_s\|_2}
\end{equation*}

\State PC function's estimation:
%\begin{equation*}
%\vect{f} \gets (\vect{\Psi}^T\vect{\Psi} + \lambda \vect{R}_{1} \vect{R}_{0}^{-1} \vect{R}_{1})^{-1} \vect{\Psi}^T \vect{X}^T \vect{u}
%\end{equation*}
$\vect{f}$ such that %
\begin{equation*}
	\begin{bmatrix}
		\vect{\Psi}^T\vect{\Psi}& \lambda\vect{R}_{1}\\
		\lambda\vect{R}_{1}& -\lambda\vect{R}_{0}
	\end{bmatrix}
	\begin{bmatrix}
		\vect{f}\\
		\vect{g}
	\end{bmatrix}
=
	\begin{bmatrix}
		\vect{\Psi}^T\vect{X}^T\vect{u}\\
		\vect{0}
	\end{bmatrix}
\end{equation*}
\State PC function's evaluation:
\begin{equation*}
\vect{f}_s \gets \vect{\Psi}^T \vect{f}
\end{equation*}

\State Repeat Steps 2--4 until convergence

\State Normalization:
\begin{equation*}
\hat{f}(p) \gets \frac{\vect{f}^T\vect{\psi}(p)}{\|\vect{f}^T\vect{\psi}\|_{L^2(\mathcal{M}_\mathcal{T})}}
\end{equation*}
\end{algorithmic}
\end{algorithm}

The problems (\ref{eq:model})-(\ref{eq:model_sparsedata}) are non-convex minimization problems in $(\vect{u}, f)$. However, in the previous section we proved the existence and uniqueness of the minimizing $f$ given $\vect{u}$ and vice-versa. This implies that the objective function is non-increasing under the update rules of the Algorithm~\ref{alg:fPCA}. Since the first guess of the PC function, given by the SVD, is usually a good starting point, in all our simulations no convergence problem has been detected.

\subsection{Parameters selection}\label{sec:par_selection}
The SM-FPCA model has a smoothing parameter $\lambda > 0$ that adjusts the trade-off between the fidelity of the estimate to the data, via the sum of the squared errors, and the smoothness of the solution, via the penalty term. The problem of choosing the smoothing parameter is common to all smoothing problems.

The flexibility given by the smoothing parameter can be seen as an advantageous feature; by varying the smoothing parameter the data can be explored on different scales. However, in many cases a data-driven automatic method is necessary. In the following simulations we consider two different criteria.
The first approach consists on a $K$-fold cross validation. The data matrix $\vect{X}$ is partitioned by rows into $K$ roughly equal groups. For each group of data $k=1,\ldots,K$ the dataset can be split into a validation set $\vect{X}^{k}$, composed of the elements of the $k$th group, and a training set, composed of the remaining elements. For different smoothing parameters, the loading function $f^{-k}$ is estimated from the training dataset. Given the estimated loading function $f^{-k}$, the associated score vector $\vect{u}^{k}$ is computed on the validation dataset. Since $f^{-k}$ has been computed on the training dataset, $\vect{u}^{k}$ should be computed on the validation dataset via the formula (\ref{eq:u_full}), where $\int_{\mathcal{M}} \! \Delta^2_{\mathcal{M}}$ can be approximated by $\vect{g}^T \vect{R}_0 \vect{g}$, being $g_h(p) = \vect{\psi}(p)^T \vect{g}$ the auxiliary function approximating $\Delta_{\mathcal{M}} f$.
Finally, we select the value of the parameter $\lambda$ that minimizes the following score:
\begin{equation}\label{eq:CV}
CV(\lambda) = \sum_{k=1}^K  \frac{\sum_{i=1}^{n} \sum_{j=1}^{s} x_i(p_j) - u_i^{k} f^{-k}(p_j))^2}{n p}.
\end{equation}

The second approach is based on the minimization of a generalized cross-validation (GCV) criteria integrated on the regression step of the iterative algorithm.
Setting $\vect{S}(\lambda) = \vect{\Psi}^T (\vect{\Psi}^T\vect{\Psi} + \lambda \vect{R}_{1} \vect{R}_{0}^{-1} \vect{R}_{1})^{-1} \vect{\Psi}^T$, the GCV score is defined as
\begin{equation*}\label{eq:GCV}
\text{GCV}(\lambda) = \frac{1}{s} \frac{\|(\vect{I} - \vect{S}(\lambda))(\vect{X}^T\vect{u})\|^2}{(1 - \frac{1}{s}
tr \{\vect{S}(\lambda)\})^2}.
\end{equation*}
The GCV score represents the average misfit of the regression model with a leave-one-out cross-validation strategy on the observations' vector $\vect{X}^T \vect{u}$. However, excluding the $i$th element from the vector $\vect{X}^T \vect{u}$ can be interpreted as removing $i$th column from the data-matrix $\vect{X}$. Thus, in terms of the data-matrix, this strategy can be interpreted as a leave-one-column-out cross-validation strategy, as opposed to the $K$-fold, where the data matrix $\vect{X}$ is partitioned by rows. The GCV approach is generally faster then the $K$-fold approach. However, $K$-fold does not require the inversion of any matrix. This is an advantageous feature, since generally the inverse of sparse matrix is not sparse. It is thus applicable also to datasets $\vect{X}$ with a large number of columns $s$.%
%On the other hand, the GCV approach can be applied to situations where only the covariance matrix is available, since the two steps algorithm can be reduced to the single iterative step
%\begin{equation}
%\vect{f}^{(new)} \gets (\vect{\Psi}^T\vect{\Psi} + \lambda \vect{R}_{1} \vect{R}_{0}^{-1} \vect{R}_{1})^{-1} \frac{\vect{\Psi}^T \vect{X}^T \vect{X}\vect{f}^{(old)}_s}{\|\vect{X}\vect{f}^{(old)}_s\|},
%\end{equation}
%depending only on the matrix $\vect{X}^T\vect{X}$, proportional to the covariance matrix. --Just for the first- Se SPCA for a more complete discussion -

\subsection{Total explained variance}
Another parameter that must be chosen is the number of PCs that satisfactorily reduces the dimension of the data. A classical approach consists on selecting this parameter on the basis of cumulated explained variance of the PC. While in the ordinary PC, the scores vectors are uncorrelated and their loadings are orthogonal, in our formulation neither the loadings are explicitly imposed to be orthogonal nor the PC scores to be uncorrelated. It is nevertheless possible to define an index of explained variance as follows. Let $\vect{\hat{U}}$ be the $n \times k$ matrix such that the columns of $\vect{\hat{U}}$ are the first $k$ PC scores vectors. Since in our estimation procedure the PC scores are normalized to have unitary norm, the variance of the PCs is captured by the PC functions. It is thus necessary to consider here the unnormalized PC scores, obtained by multiplying each score vector by the norm of the associated PC function. Without the uncorrelation assumption, it is meaningless to compute the total variance explained by the first $k$ PCs by tr$(\vect{\hat{U}}^T\vect{\hat{U}})$. To overcome this problem \cite{zou2004} propose to remove linear dependence between correlated PC scores vectors, by regression projection. Thus they compute the QR decomposition of $\vect{\hat{U}}$ as $\vect{\hat{U}} = \vect{Q}\vect{R}$ and define the \textit{adjusted total variance} as $\sum_{j=1}^k \vect{R}_{jj}^2$, where $\vect{R}_{jj}$ represents the variance explained by the $j$th PC that is not already explained by the previous $j-1$ components.

\section{Simulation studies}\label{sec:simulations}
In this section we conduct simulations to assess the performance of the SM-FPCA algorithm compared to other methods.

\begin{figure}[h] % figuur 1
%\vspace{6pc}
\includegraphics[width=\textwidth]{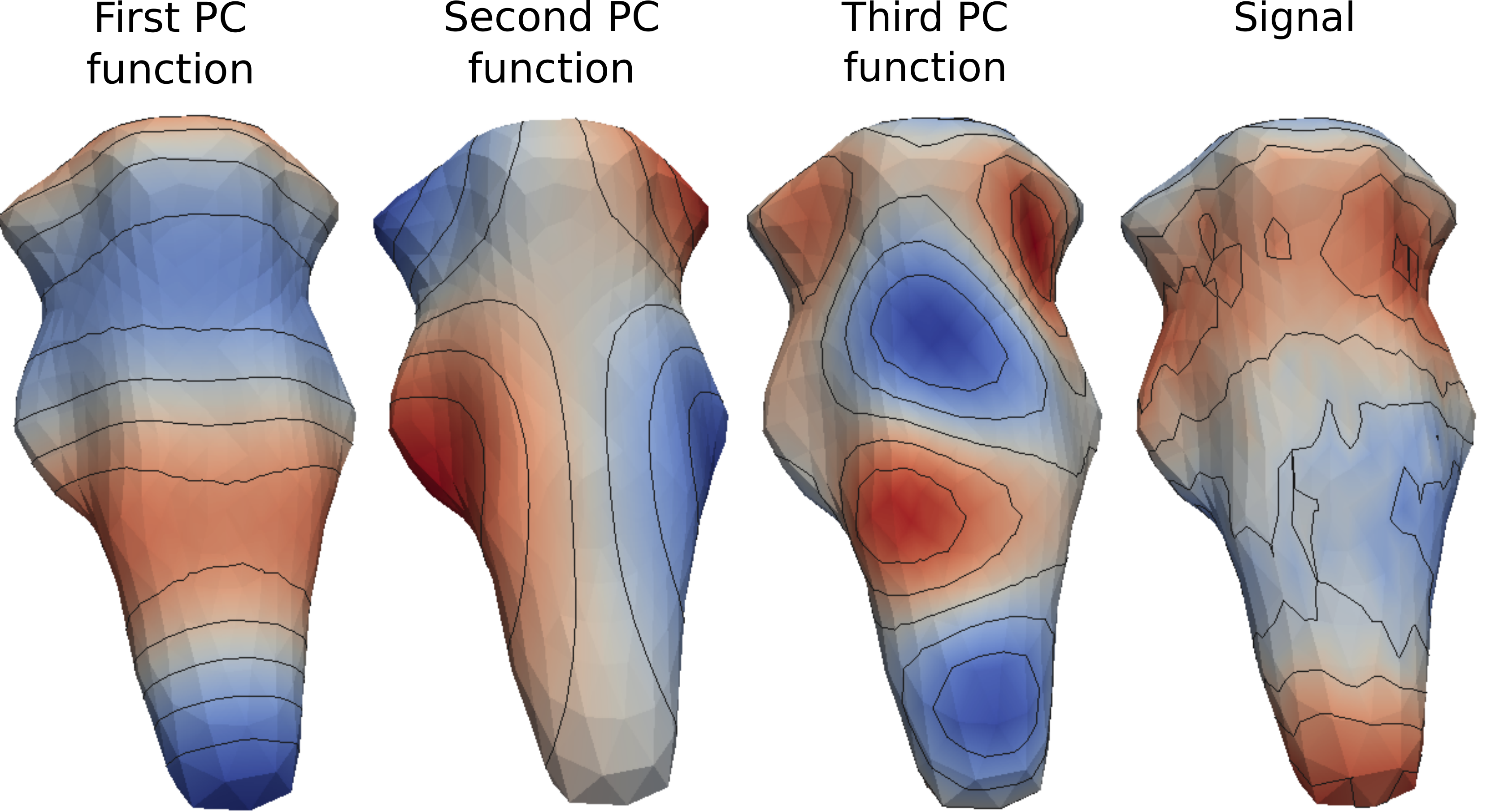}
\caption[]{From left to right, a plot of the true first, second and third PC functions and a plot of a noisy observation on the brainstem, generated from these three PC functions.}
\label{fig:brain_stem}
\end{figure}

We consider as domain of the functional observations a triangulated surface $\mathcal{M}_\mathcal{T}$ with 642 nodes that approximates the brainstem. On this triangulated surface we generate the orthonormal functions $\{v_l\}_{l=1,2,3}$, consisting in three eigenfunctions of the Laplace-Beltrami operator, as shown in Figure~\ref{fig:brain_stem}. These functions represent the first three PC functions. We then generate $n = 50$ smooth functions $x_1,\ldots,x_{50}$ on $\mathcal{M}_\mathcal{T}$ by
\begin{equation}
x_i = u_{i1} v_1 + u_{i2} v_2 + u_{i3} v_3 \quad i=1,\ldots,n,
\end{equation}
where $u_{i1}$, $u_{i2}$, $u_{i3}$ are independent random variables that represent the scores and are distributed as $u_{il} \sim \mathcal{N}(0,\sigma_l^2)$, with $\sigma_1 = 5$, $\sigma_2 = 3$ and $\sigma_3 = 1$.
The smooth functions $x_i$ are then sampled at locations $p_j \in \mathbb{R}^3$ with $j = 1,\ldots, s$ coinciding with the nodes of the triangulates surface. Moreover at each of these points we add to the functions a Gaussian noise with mean zero and standard deviation $\sigma = 0.1$ to obtain the noisy observations denoted with $x_i(p_j)$. We are thus interested in recovering the smooth PC functions $\{v_l\}_{l=1,2,3}$ from these noisy observations over $\mathcal{M}_\mathcal{T}$. We compare the proposed SM-FPCA technique to two alternative approaches.

The first basic approach we consider is a simple multivariate PCA (MV-PCA) applied to the data-matrix $\vect{X}$. The PC functions are thus obtained by piecewise linear interpolation over the mesh $\mathcal{M}_{\mathcal{T}}$. Finally they are normalized to have unitary norm in $L^2(\mathcal{M}_{\mathcal{T}})$.

A second natural approach is based on a pre-smoothing of the noisy observations that tries to recover the smooth functions $x_i, \, i=1,\ldots,n$, from their noisy observations $x_i(p_j)$, followed by a MV-PCA on the denoised evaluations of the functions on $p_j, \, j=1,\ldots,s$. The smoothing problem for a field defined on a Riemannian manifold is not trivial. In this case the smoothing technique applied is Iterated Heat Kernel (IHK) smoothing [\cite{chung2005}]. The heat kernel smoothing of the noisy observation $x_i(p_j)$, is given by
$K_\eta \times x_i(p_j) = \int_{\mathcal{M}} K_\eta (p,q) x_i(p_j) dq$, where $\eta$ is the smoothing parameter and $K_\eta$ is the heat kernel, whose analytic expression can be extracted from the eigenfunctions of the Laplace-Beltrami operator. However, for numerical approximation, it can be shown that for $\eta$ small and for $q$ close to $p$ we have
\begin{equation*}
K_\eta (p,q) \approx \frac{1}{(2 \pi \eta)^{\frac{1}{2}}} exp[-\frac{d^2(p,q)}{2 \eta^2}].
\end{equation*}
The desired level of smoothing can be reached after $k$ iterations, thanks to the following property: $K_\eta^{k} \times f = K_\eta \times \ldots \times K_\eta \times f = K_{\sqrt{k}\eta}$. For a fixed bandwidth $\eta$, the level of smoothing is determined by an optimal number of iterations selected via the F-test criterion outlined in \cite{chung2005}. In these simulations, the bandwidth has been set at $\eta=2.5$, heuristically selecting the one with the best performance after some initial pilot studies. We refer to this approach as IHK-PCA.

The proposed SM-FPCA technique is implemented as follows. For each PC we run Algorithm 1 with 15 iterations of the steps 2-4. For the choice of the optimal smoothing parameter $\lambda$, both $K$-fold, with $K = 5$, and GCV approaches have been applied.

\begin{figure}[!htb] % figuur 1
%\vspace{6pc}
\includegraphics[width=1\textwidth]{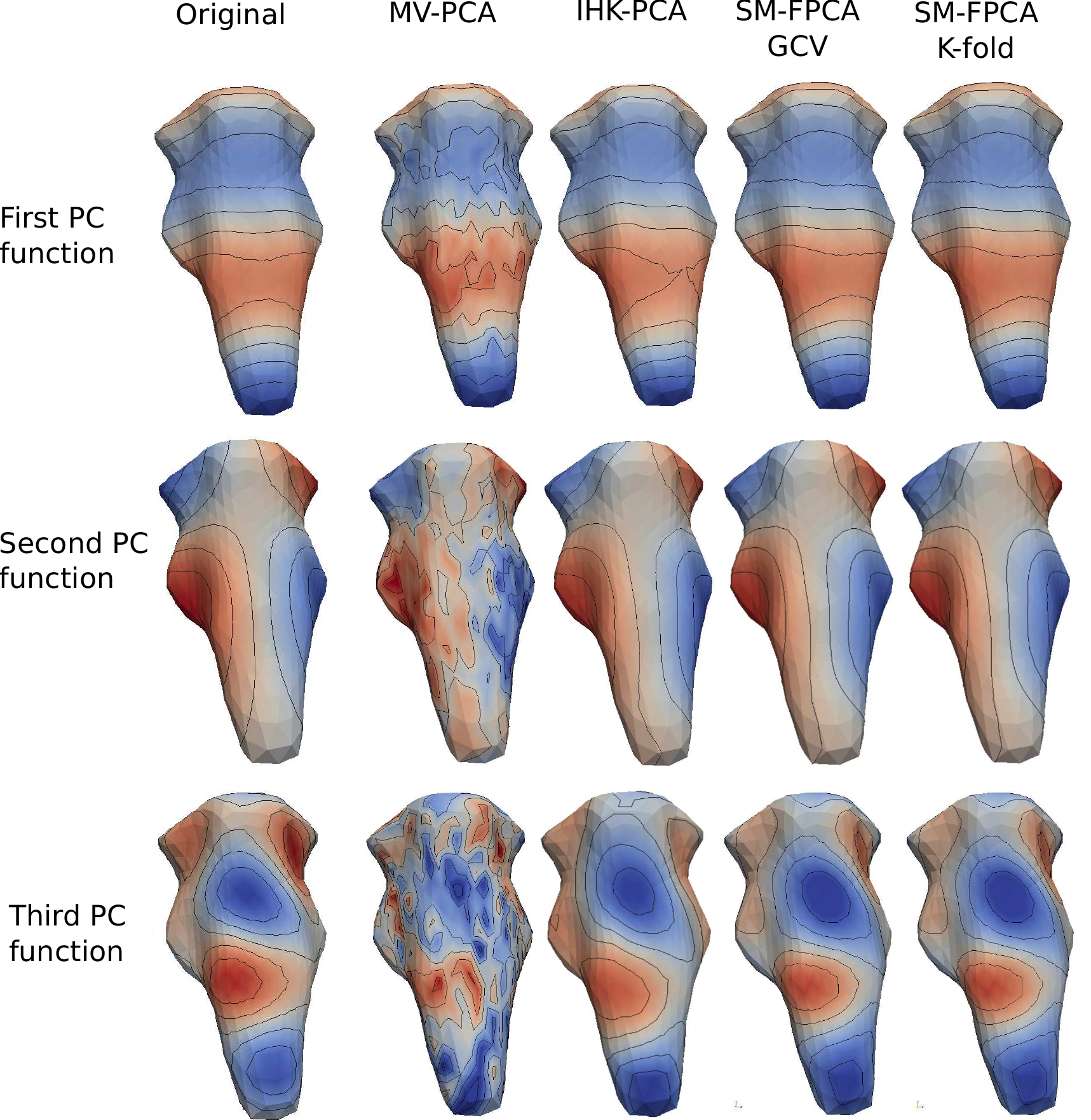}
\caption[]{From left to right, contours of the original PC functions and their estimates respectively with MV-PCA, IHK-PCA, SM-FPCA GCV and SM-FPCA K-fold. From a visual inspection, MV-PCA shows unsatisfactory results, while a better estimation is achieved by IHK-PCA and SM-FPCA. In particular SM-FPCA is able to better capture details that IHK-PCA ignores. This is apparent for instance in the third PC function reconstruction, in the top-left and top-right corners.}
\label{fig:comparison}
\end{figure}

\begin{figure}[!htb] % figuur 1
%\vspace{6pc}
\includegraphics[width=1\textwidth]{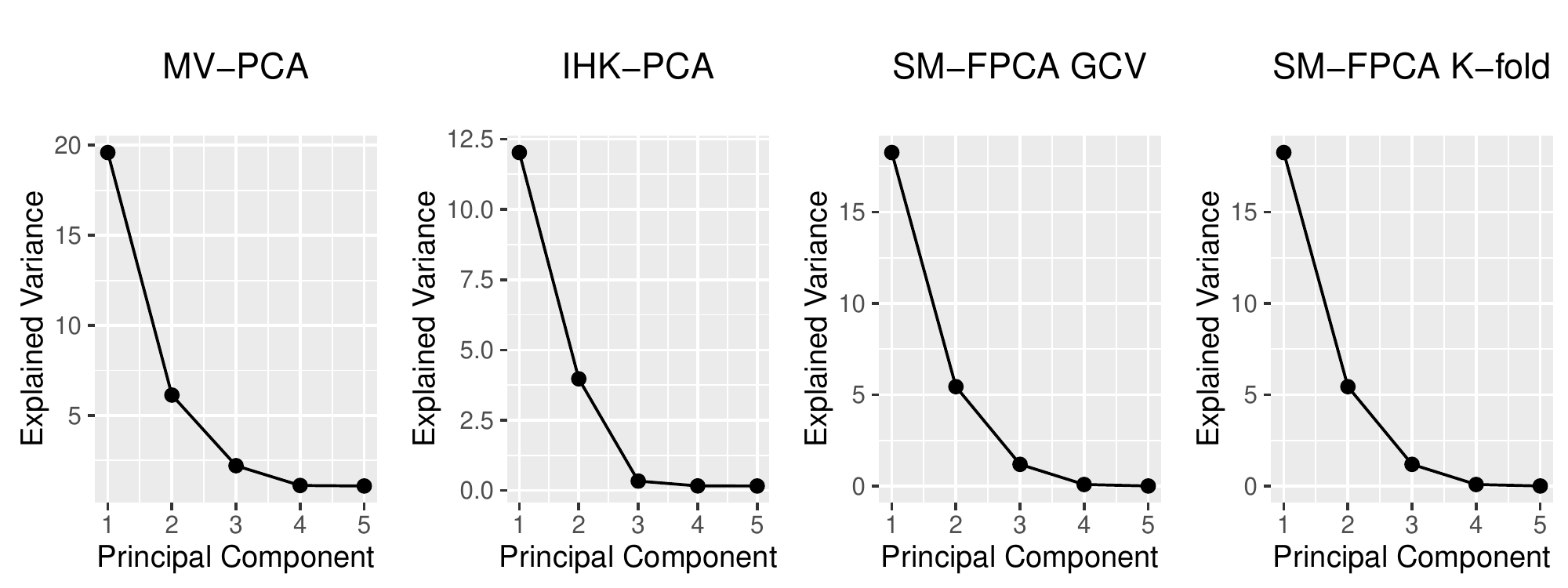}
\caption[]{From left to right, plot of the empirical variances explained by the first 5 PCs computed with MV-PCA, IHK-PCA, SM-FPCA GCV and SM-FPCA K-fold.}
\label{fig:cumplot}
\end{figure}

The reconstructed PC functions, using the three different approaches are shown in Figure~\ref{fig:comparison}. It is evident that applying the MV-PCA yields to a reconstruction far from the true, because of the absence of any spatial information. The reconstruction through the IHK-PCA approach and the \mbox{SM-FPCA} model are considerably more satisfactory. In Figure~\ref{fig:cumplot} we show the plots with the cumulative percentage of explained variance, where in the case of SM-FPCA, the explained variance has been computed as detailed in the Section~\ref{sec:par_selection}.

\begin{figure}[!htb] % figuur 1
%\vspace{6pc}
\includegraphics[width=\textwidth]{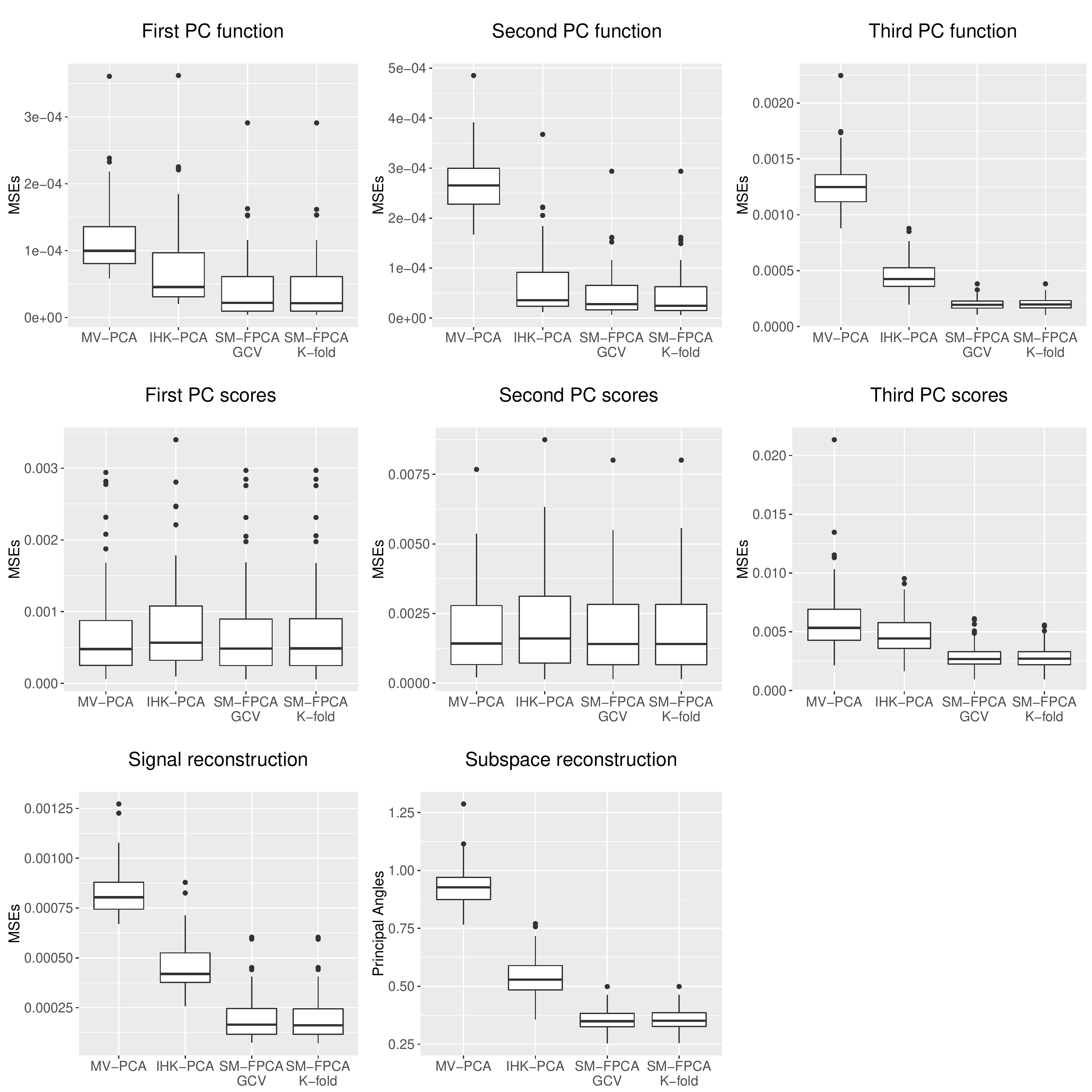}
\caption[]{Boxplots summarizing the performance of IHK-PCA and SM-FPCA. For the SM-FPCA both GCV and $K$-fold have been applied for the selection of the smoothing parameter.}
\label{fig:boxplots}
\end{figure}

While the poor performance of the MV-PCA is evident, to assess the performance of the other two methods, we apply them to 100 datasets generated as previously detailed. The quality of estimated individual surfaces is then measured using the mean square error (MSE) over all the locations $p_j, \, j=1,\ldots,s$. MSEs are also used to evaluate the reconstruction of the PC scores vectors.  Another performance measure used is the principal angle between the subspace spanned by the estimated PC functions and the subspace spanned by the true PC functions, as used in \cite{shen2008}. Intuitively, the principal angle measures how similar the two subspaces are. For this purpose we construct the $s \times 3$ matrices $\mathbb{V} = (v_i(p_j))$ and $\mathbb{\hat{V}} = (\hat{v}_i(p_j))$, where $\hat{v}_i$ is the $i$th estimate of the true PC function $v_i$. Then we compute the orthonormal set of basis $\vect{Q}_\mathbb{V}$ and $\vect{Q}_\mathbb{\hat{V}}$ from the QR decomposition of $\mathbb{V}$ and $\mathbb{\hat{V}}$. The principal angle is defined as the angle $cos^{-1}(\rho)$, where $\rho$ is the minimum singular value of $\vect{Q}^T_\mathbb{\hat{V}}\vect{Q}_\mathbb{V}$. The results are summarized in the boxplots in Figure~\ref{fig:boxplots}, which compares the MV-PCA, IHK-PCA and SM-FPCA algorithms with respect to the reconstruction's errors of the PC functions $\{v_l\}_{l=1,2,3}$, the PC scores $\{\vect{u}_l \}_{l=1,2,3}$ where $\vect{u}_l = (u_{il})$, the reconstructed signals $x_i = u_{i1} v_1 + u_{i2} v_2 + u_{i3} v_3$ for $i=1,\ldots,50$ and the principal angles between the subspaces spanned by the true and estimated PC functions.

The boxplots highlight the fact that SM-FPCA provides the best estimates of the PC functions, corresponding scores vectors, signals and subspace reconstruction.

%\begin{figure}[h] % figuur 1
%%\vspace{6pc}
%\includegraphics[width=0.7\textwidth]{Figures/contours.jpg}
%\caption[]{From left to right, contours of the original first weight function and of 10 reconstruction with IHK-PCA and Manifold-fPCA}
%\label{fig:PC_func}
%\end{figure}

\section{Application}\label{sec:app}
The data set which we consider in this paper arises from the Human Connectome Project Consortium [HCP, \cite{vanessen2012}], which is collecting data such as structural scans, resting-state and task-based functional MRI scans, and diffusion-weighted MRI scans from a large number of healthy volunteers to help elucidate normal brain function.
Many preprocessing considerations have already been resolved in the so-called minimally preprocessed dataset. Among the various preprocessing pipelines applied to the HCP original data, of particular interest for us is the one named \textit{fMRISurface} [\cite{glasser2013}]. This pipeline provides a transformation of the 3D structural MRI and 4D signal from the functional MRI scan, so to enable the application of statistical analysis techniques on brain surfaces. For each subject, the personal cortical surface is extracted as a triangulated surface from the structural MRI and to each vertex of this mesh is associated a BOLD time-series derived from the BOLD signal of the underlying gray-matter ribbon. The extracted cortical surfaces are aligned to a template cortical surface generated from the cortical surfaces of 69 healthy adults. In practice, this cortical surface is represented by two triangulated surfaces with 32k vertices, one for each hemisphere. In Figure~\ref{fig:brain_mesh} the left hemisphere is shown. Through this anatomical transformation map, the patients' BOLD time-series, on the cortical surface, are coherently located to the vertices of the template cortical surface. This, of course, raises questions about the implications of anatomical alignment, and a small simulation study in the supplementary material investigates this issue. The fMRI signal used for our analysis has been acquired in absence of any task and for this reason is also called resting state fMRI. Finally each time-series is filtered to the band of frequencies $[0.009,0.08]$Hz. Summarizing, the data considered are fMRI filtered time-series on a common triangulated template mesh.

%We wish to combine the classical approaches used to analyse rsfMRI data and the surface-constrained PCA algorithm implemented in this work, to allow PCA based rsfMRI analysis directly on the cortical surface.
As already mentioned in Section~\ref{sec:intro}, a classic approach in the study of the resting state fMRI is to exploit the time dimension of the data, for the extraction of a connectivity measure among the different parts of the cortical surface. A standard choice for this purpose is the computation of the temporal correlation. It first consists of identifying a Region of Interest (ROI) on the cortical surface. This is the area whose behaviour, as compared to the rest of the cortical surface, is of interest for the investigator. Within each subject, a cross-sectional average of all the time-series in the ROI is used to find a representative mean time-series. To each vertex of the cortical surface we associate the pairwise correlation of the time-series located in that vertex with the subject-specific time-series representative of the ROI. Finally each correlation value is transformed using Fisher's r-to-z transformation, yielding a resting state functional connectivity (RSFC) map for each subject. The total number of subjects considered for this analysis is 491.

\begin{figure}[h]
\centering
\begin{subfigure}{.5\textwidth}
  \centering
  \includegraphics[width=0.85\textwidth]{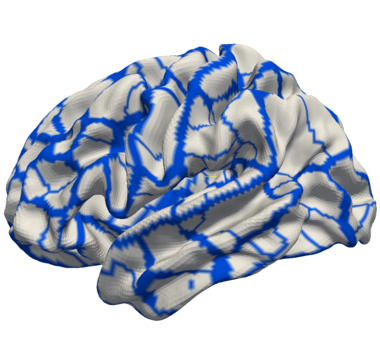}
\end{subfigure}%
\begin{subfigure}{.5\textwidth}
  \centering
  \includegraphics[width=0.85\textwidth]{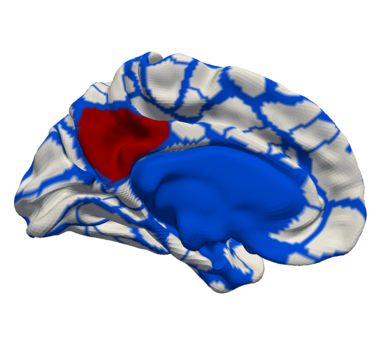}
\end{subfigure}
\caption{Parcellation of the cortical surface derived in \cite{gordon2014}. In red the Region of Interest chosen for the computation of the RSFC maps. This region is localized on an area of the cerebral cortex called precuneus. The blue colours indicate the parcellated regions, with the major blue area being the join between the two brain hemispheres, which does not lie on the manifold surface and which is therefore excluded from the cortical surface analysis.}
\label{fig:parcellation}
\end{figure}

\begin{figure}[h]
\centering
\begin{subfigure}{.5\textwidth}
  \centering
  \includegraphics[width=1\textwidth]{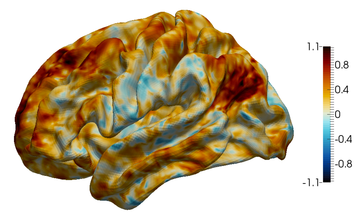}
\end{subfigure}%
\begin{subfigure}{.5\textwidth}
  \centering
  \includegraphics[width=1\textwidth]{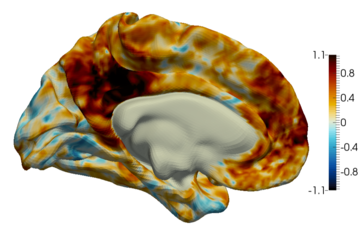}
\end{subfigure}
\caption{A snapshot of the RSFC map of one subject.}
\label{fig:RSFC}
\end{figure}

For the choice of the ROI, we consider the cortical parcellation derived in \cite{gordon2014}, where a group-average boundary map of the cortical surface is derived from resting state fMRI (Figure~\ref{fig:parcellation}). The identified cortical areas are unlikely to correspond the individual parcellation of each subject, since they are derived from a group average study. However, they can serve as a reasonable ROIs in individual subjects. The parcel that served as ROI in the following analysis is highlighted in red in Figure~\ref{fig:parcellation}.
For the chosen ROI, a snapshot of the RSFC map of one subject is shown in Figure~\ref{fig:RSFC}.
%\vspace{-10pt}

\begin{figure}[h]
\centering
\begin{subfigure}{.5\textwidth}
  \centering
  \includegraphics[width=1\textwidth]{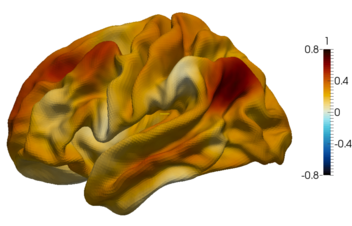}
\end{subfigure}%
\begin{subfigure}{.5\textwidth}
  \centering
  \includegraphics[width=1\textwidth]{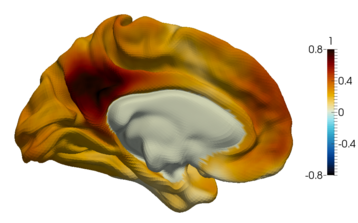}
\end{subfigure}
\caption{The mean RSFC map computed over 491 subject. As expected, high correlation values are visible inside the ROI.}
\label{fig:mean}
\end{figure}

The mean RSFC map is shown in Figure~\ref{fig:mean}. As expected high correlation values are visible inside the ROI. The mean RSFC over 491 subjects shows a variability coherent with the parcellation, in the sense that the vertices inside each parcel show similar values. We wish now to understand which are the main modes of variation of these RSFC maps among the different subjects, by applying a PCA.

\begin{figure}[!htb]
\centering
\begin{subfigure}{\textwidth}
  \centering
  \includegraphics[width=\textwidth]{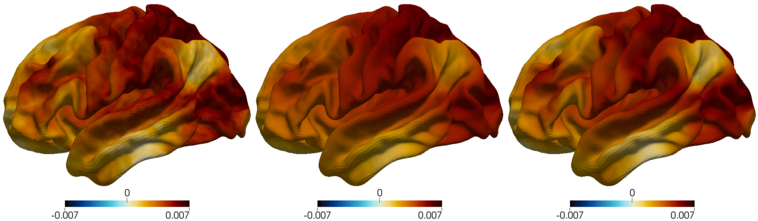}
\end{subfigure}
\begin{subfigure}{\textwidth}
  \centering
  \includegraphics[width=\textwidth]{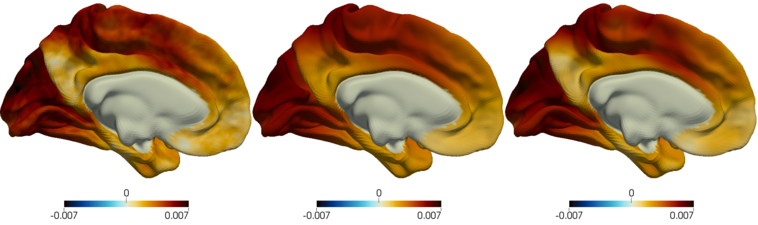}
\end{subfigure}
\caption{From left to right, two views of the first PC function computed respectively with MV-PCA, IHK-PCA and SM-FPCA.}
\label{fig:PC1_brain}
\end{figure}

\begin{figure}[!htb]
\centering
\begin{subfigure}{\textwidth}
  \centering
  \includegraphics[width=1\textwidth]{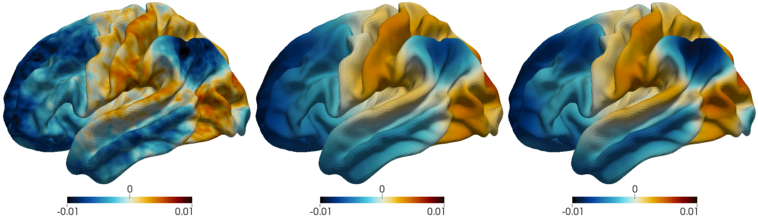}
\end{subfigure}
\begin{subfigure}{\textwidth}
  \centering
  \includegraphics[width=1\textwidth]{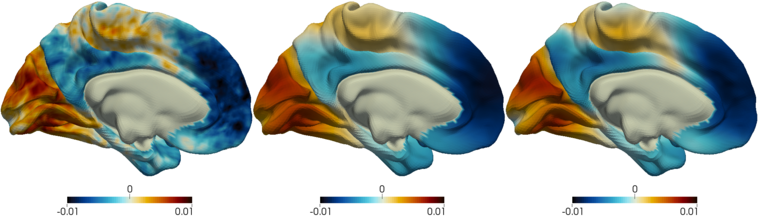}
\end{subfigure}
\caption{From left to right, two views of the second PC function computed respectively with MV-PCA, IHK-PCA and SM-FPCA.}
\label{fig:PC2_brain}
\end{figure}

\begin{figure}[!htb]
\centering
\begin{subfigure}{\textwidth}
  \centering
  \includegraphics[width=1\textwidth]{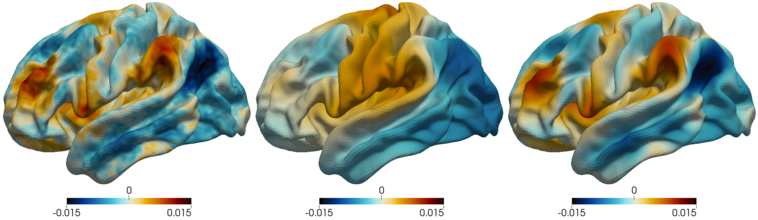}
\end{subfigure}
\begin{subfigure}{\textwidth}
  \centering
  \includegraphics[width=1\textwidth]{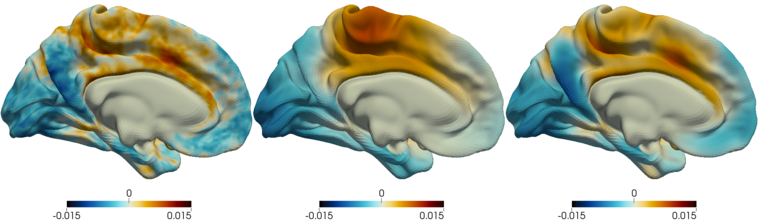}
\end{subfigure}
\caption{From left to right, two views of the third PC function computed respectively with MV-PCA, IHK-PCA and SM-FPCA.}
\label{fig:PC3_brain}
\end{figure}

%However the application of a MV-PCA to estimate the PC functions from the RSFC maps gives unsatisfactory results.  %
The first three PC functions, estimated with SM-FPCA, are shown in Figures~\ref{fig:PC1_brain}-\ref{fig:PC2_brain}-\ref{fig:PC3_brain} as compared to the PC functions derived from MV-PCA and IHK-PCA. The choice of the smoothing parameter for the SM-FPCA is based on the $K$-fold cross validation, with $K = 5$.

The PC functions estimated from the MV-PCA shows an excessive variability, since the sample size is not sufficiently large to deal with the extremely high dimensionality of the data, and the spatial information is completely ignored by this model. In fact, even recent attempts to model the subject variability from resting state fMRI leads to the conclusion that spatial mismatches, introduce by the alignment problem, are one of the biggest sources of currently observable differences between subjects [\cite{harrison2015}]. This registration process can result in misalignments, due to the lack to functional regions being perfectly coincident or due to situations where the local topology is strongly different among subjects. These misalignments can introduce fictitious effects on the computed PC functions. Data misalignment is a well known problem in FDA [\cite{marron2015}]. For functional data with one-dimensional domains, typical approaches are based on shifting or (monotone) transformations of the domain of each function. But neither shifting nor monotonic transformations make sense on a generic non-Euclidean domain, so it is not clear how to generalize the standard FDA approaches. The introduction of a smoothing penalty in the PCA model should reduce the variability effects due to misalignment. In fact the smoothing parameter in the SM-FPCA algorithm can be seen as a further degree of freedom that allows a multiscale analysis, meaning that by increasing the smoothing penalty parameter is possible to constrain the results to show only the macroscopical effects of the phenomena and to remove the artifacts introduced by the preprocessing steps.

%The PC functions estimated through IHK-PCA and SM-FPCA seem both to resolve the problem of the presence of some residual noise in the PC function estimates. However, we would like to emphasize the fact that, for instance, the third PC function computed with IHK-PCA shows some differences with the one computed with SM-FPCA. This differences may be due to the fact that an individual pre-smoothing approach with the presented data tends to delete part of the information contained in the single RSFC map, while the SM-FPCA model incorporates all the noisy data as is and introduces a smoothness penalization directly on the PC function estimates. Contrary to MV-PCA and IHK-PCA, in all three PC functions, SM-FPCA shows a satisfactory level of smoothness, without deleting sharper changes in some locations, which may indicate the presence of a boundary between two cortical areas, since it is well known that the cortical surface is organized in several interacting cortical areas.
Both IHK-PCA and SM-FPCA returns smooth PC functions. A visual inspection of the estimated PC functions though highlights that IHK-PCA completely smooth out sharper changes in the modes of variations, missing some localized features that are apparent in MV-PCA and are also very well captured by the proposed SM-FPCA. Comparing for instance the estimated third PC functions, in the top views of Figure~\ref{fig:PC3_brain}, one can see for both MV-PCA and SM-PCA corresponding localized areas with very high values (in red) and very low values (in blue) that are instead missing in the IHK-PCA estimate. By contrary, the pre-smoothing approach appears to introduce some artifacts: looking at the bottom views in Figure~\ref{fig:PC3_brain}, one can for instance notice that IHK-PCA estimated third PC function has high values in the higher part of the plot, that do not have match neither on the MV-PCA nor on the SM-FPCA estimate.

\begin{figure}[H]
\centering
\begin{subfigure}{.33\textwidth}
  \centering
  \includegraphics[width=1\textwidth]{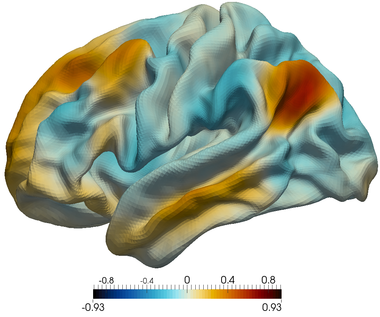}
\end{subfigure}%
\begin{subfigure}{.33\textwidth}
  \centering
  \includegraphics[width=1\textwidth]{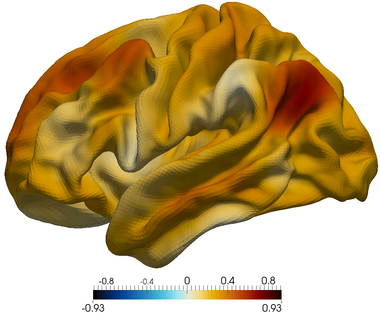}
\end{subfigure}%
\begin{subfigure}{.33\textwidth}
  \centering
  \includegraphics[width=1\textwidth]{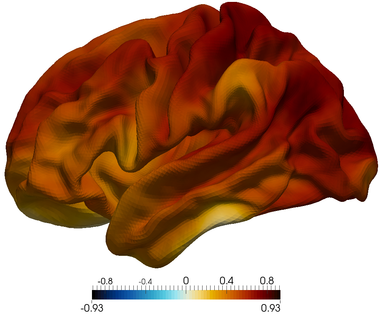}
\end{subfigure}
\begin{subfigure}{.33\textwidth}
  \centering
  \includegraphics[width=1\textwidth]{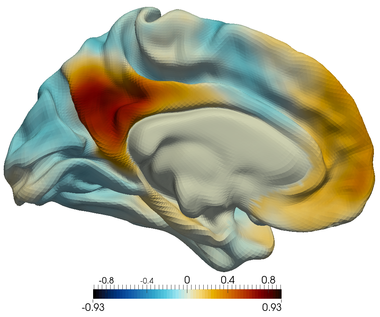}
\end{subfigure}%
\begin{subfigure}{.33\textwidth}
  \centering
  \includegraphics[width=1\textwidth]{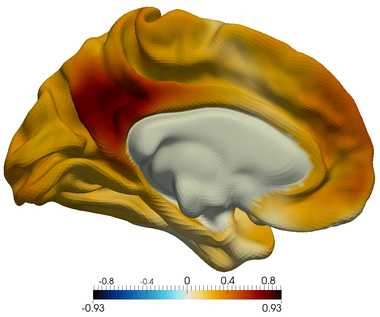}
\end{subfigure}%
\begin{subfigure}{.33\textwidth}
  \centering
  \includegraphics[width=1\textwidth]{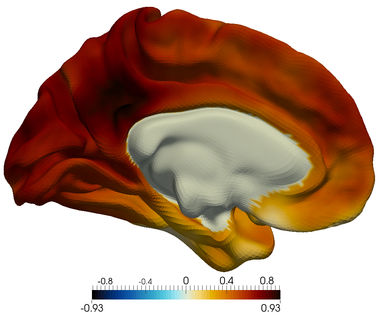}
\end{subfigure}
\caption{From left to right, two views of $\mu - 2\sigma f$, $\mu$, $\mu + 2\sigma f$, where $\mu$ denotes the mean RSFC map, $\sigma$ denotes the standard deviation of the first PC scores vector and $f$ denotes the first PC function.}
\label{fig:PC1_interpretation}
\end{figure}

For the purpose of interpretation of the PC functions, we might prefer to plot the functions $\mu \pm 2\sigma f$, where $\mu$ denotes the mean RSFC map, $\sigma$ denotes the standard deviation of the PC scores vector and $f$ denotes the associated PC function. In Figure~\ref{fig:PC1_interpretation} we show the described plot for the first PC function. We can observe that while the high correlation value in the ROI and inferior parietal are in first approximation preserved from subject to subject, a high variability between subjects can be observed in the areas surrounding the ROI and the inferior parietal, which is understood due to individual inter-subject differences [\cite{buckner2008brain} and references therein]. However, it should be noted that variability can be both somewhat localised as well as more spatially smooth, indicating that even in resting state data, brain regions have differential response which is not simply a result of noise in the data.

\section{Discussion}\label{sec:discussion}
In this paper we introduced a novel PCA technique that can handle functional data located over a two-dimensional manifold. The adopted approach is based on a regularized PCA model. In particular, a smoothness penalty term that measures the curvature of a function over a manifold is considered and the estimation problem is solved via an iterative algorithm that uses finite elements. The motivating application is the analysis the RSFC maps over the cortical surface, derived from fMRI.  In this setting the adoption of a MV-PCA suffers of the high-dimensionality of the data with respect to the relatively small sample size. The adoption of an approach based on individual pre-smoothing of the functional samples, followed by a MV-PCA, gives smooth estimates of the PC functions. However, this pre-smoothing step tends to remove useful information from the original data. The proposed SM-FPCA instead returns smooth PC functions that nevertheless are able to capture localized features of the estimated PC functions. It could also be imagined that in more complex study designs (such as patient versus control studies) these PC functions, along with the associated scores, could be used to investigate diverse difference between groups or covariate effects.

A further important feature of SM-FPCA is its computational efficiency. The most computationally intensive operation is the resolution of the linear system in the iterative algorithm. However this linear system enjoys two important properties. The first is the independence between its dimensions, related to the number of nodes of the triangular mesh, and the number of point-wise observations available for each functional sample as well as the sample size. In fact, since its resolution time depends mostly on the mesh size, a mesh simplification approach [\cite{SSRM2}] could be adopted to speed up the algorithm. The second and most fundamental property is the sparsity of the linear system. The use of a sparse solver allows an efficient computation of the solution. For instance, in the final application the dimension of the linear system is $64$K$\times64$K. Despite its dimension, the solving time is less than a second. The application of the entire algorithm, for a fixed smoothing parameter, with 15 iterations is less than 15 seconds on a Intel Core i5-3470 3.20GHz workstation, with 4 GB of RAM.

\appendix
\appendixpage

\section{Surface Finite Element Discretization}\label{app1}

%We approach the minimization of the equations (3.1) in the original paper through an iterative algorithm in $\vect{u}$ and $f$. It is trivial to show that the minimization of the objective function, in the two steps of the iterative algorithm, reduces to the following equations:
%\begin{enumerate}[label=\textit{Step} \arabic*,align=left, leftmargin=1.0cm]
%\step Estimation of the unitary-norm vector $\vect{u}$ for a fixed $f$.
%\begin{equation*}\label{eq:u}
%\vect{u} = \frac{\vect{X}\vect{f}_s}{\|\vect{X}\vect{f}_s\|_2}.
%\end{equation*}
%\step Estimation of $f$ for a fixed $\vect{u}$. $f$ is such that it minimizes
%\begin{equation*}
%J_{\lambda, \vect{u}}(f) = \vect{f}_s^T\vect{f}_s+\lambda \int_{\mathcal{M}} \! \Delta^2_{\mathcal{M}} f -2\vect{f}_s^T\vect{X}^T\vect{u}.
%\end{equation*}
%\end{enumerate}
%Now we give the details of the discretization procedure that brings from the infinite-dimensional problem of estimating $f \in H^2(\mathcal{M})$ for a fixed $\vect{u}$, to its discrete formulation.

\subsection{Well-posedness of the estimation problem (\ref{eq:problem})}
\begin{proof}{Proposition 1.}
We exploit a characterization theorem [\cite{Breass}, chapter 2] which states that if $G$ is a symmetric, positive definite, bilinear form on a vector space $L$, and $F$ is a linear functional on $L$, then $v$ is the unique minimizer of
\begin{equation*}
G(v,v) - 2F(v)
\end{equation*}
in $V$ if and only if
\begin{equation}\label{eq:var_app}
G(v,\varphi) = F(\varphi) \qquad \text{ for all } \varphi \in L.
\end{equation}
Moreover, there is at most one solution to problem \ref{eq:var_app}.

The desired result follows from application of the above theorem considering the vector space $L = H^2(\mathcal{M})$, the symmetric, positive definite, bilinear form $G(f,\varphi):= \sum_{j=1}^p\varphi(p_j)f(p_j) + \lambda \int_\mathcal{M} \Delta\varphi \Delta f$ and the linear functional\\ $F(f)=\sum_{j=1}^p f(p_j) \sum_{i=1}^n x_i(p_j)u_i$.
Positive definitiveness of the form $G$, in $ H^2(\mathcal{M})$, is shown by the following argument.
Suppose that $G(f,f)=0$ for some $f \in H^2(\mathcal{M})$; then $\int_\mathcal{M} \Delta_\mathcal{M}^2 f=0$ and $\sum_{j=1}^p f(p_j)^2=0$. Each element $f \in H^2(\mathcal{M})$ can be written such that, for any $p \in \mathcal{M}$, $f(p) = \tilde{f}(p) + c$, with $\tilde{f} \in U =\{ \tilde{f} \in H^2(\mathcal{M}): \int_\mathcal{M} \tilde{f} = 0 \}$ and $c$ a constant. The solution of $\Delta_\mathcal{M} \tilde{f} = 0$ in $U$ exists unique and is $\tilde{f}=0$ [\cite{dziuk2013}]. Thus $\int_{\mathcal{M}} \Delta_\mathcal{M}^2 f=0$ for $f \in H^2(\mathcal{M})$ implies that $f(p)=c$, for any $p \in \mathcal{M}$, then $\sum_{j=1}^p f(p_j)^2 = p c^2$. But $pc^2=0$ if and only if $c=0$, so $f(\cdot)=0$. Consequently, $G$ is positive definite on $H^2(\mathcal{M})$.

The estimator $\hat{f}$ is thus
\begin{equation} \label{eq:eulero-lagrange}
\sum_{j=1}^p \varphi(p_j)\hat{f}(p_j) + \lambda \int_\mathcal{M} \Delta_\mathcal{M} \varphi \Delta_\mathcal{M} \hat{f} = \sum_{j=1}^p \varphi(p_j)\sum_{i=1}^n x_i(p_j)u_i
\end{equation}
for every $\varphi \in H^2(\mathcal{M})$.

\end{proof}

\subsection{Reformulation of the estimation problem}

The problem of finding $f \in H^2(\mathcal{M})$ that satisfies condition (\ref{eq:eulero-lagrange}) for every $\varphi \in H^2(\mathcal{M})$ can be rewritten as the problem of finding $(\hat{f},g) \in H^2(\mathcal{M}) \times L^2(\mathcal{M})$ that satisfies:
\begin{align}\label{eq:eulero-double}
\begin{cases}
\sum_{j=1}^p \varphi(p_j)\hat{f}(p_j) + \lambda \int_\mathcal{M} (\Delta \varphi) g = \sum_{j=1}^p \varphi(p_j)\sum_{i=1}^n x_i(p_j)u_i\\
\int_\mathcal{M} vg - \int_\mathcal{M} v (\Delta \hat{f}) = 0
\end{cases}
\end{align}
for all $(\varphi,v) \in H^2(\mathcal{M}) \times L^2(\mathcal{M})$. In fact, if the pair of functions $(\hat{f},g) \in H^2(\mathcal{M}) \times L^2(\mathcal{M})$ satisfies condition (\ref{eq:eulero-double}) for all $(\varphi,v) \in H^2(\mathcal{M}) \times L^2(\mathcal{M})$, then $\hat{f}$ also satisfies problem (\ref{eq:eulero-lagrange}). In contrast, if $\hat{f} \in H^2(\mathcal{M})$ satisfies problem (\ref{eq:eulero-lagrange}), then the pair $(\hat{f},\Delta \hat{f})$ automatically satisfies the two equations in problem (\ref{eq:eulero-double}).
Owing to integration by part and to the fact that $\mathcal{M}$ has no boundaries, we get:
\begin{eqnarray*}
\int_\mathcal{M} (\Delta_\mathcal{M} \varphi) g = - \int_\mathcal{M} \nabla_\mathcal{M} \varphi \nabla_\mathcal{M} g\\
\int_\mathcal{M} v (\Delta_\mathcal{M} \hat{f}) = - \int_\mathcal{M} \nabla_\mathcal{M} v \nabla_\mathcal{M} \hat{f}
\end{eqnarray*}

Now, asking the auxiliary function $g$ and of the test functions $v$ to be such that $g, v \in H^1(\mathcal{M})$, the problem of finding $\hat{f}\in H^2(\mathcal{M})$ that satisfies (\ref{eq:eulero-lagrange}) for each $\varphi \in H^2(\mathcal{M})$ can be reformulated  as finding $(\hat{f},g) \in (H^1(\mathcal{M}) \cap C^0(\mathcal{M}) ) \times H^1(\mathcal{M})$
\begin{align}\label{eq:weak_eulero}
\begin{cases}
\sum_{j=1}^p \varphi(p_j)\hat{f}(p_j) + \lambda \int_\mathcal{M} \nabla \varphi \nabla g = \sum_{j=1}^p \varphi(p_j)\sum_{i=1}^n x_i(p_j)u_i\\
\int_\mathcal{M} vg - \int_\mathcal{M} \nabla v \nabla \hat{f} = 0
\end{cases}
\end{align}
for all $(\varphi,v) \in (H^1(\mathcal{M}) \cap C^0(\mathcal{M}) ) \times H^1(\mathcal{M})$; Moreover, the theory of problems of elliptic regularity ensure that such $\hat{f}$ still belongs to $H^2(\mathcal{M})$ [\cite{dziuk2013} and reference therein].
Finally the discrete estimators $\hat{f}_h, \hat{g}_h \in V \subset H^1(\mathcal{M})$ are obtained solving
\begin{align*}
\begin{cases}
&\! \int_{\mathcal{M}_\mathcal{T}} \nabla_{\mathcal{M}_\mathcal{T}} \hat{f}_h \nabla_{\mathcal{M}_\mathcal{T}} \varphi_h - \int_{\mathcal{M}_\mathcal{T}} \hat{g}_h \varphi_h = 0\\
&\! \lambda \!\int_{\mathcal{M}_\mathcal{T}} \nabla_{\mathcal{M}_\mathcal{T}} \hat{g}_h \nabla_{\mathcal{M}_\mathcal{T}} v_h + \sum\limits_{j=1}^s \hat{f}_h(p_j)v_h(p_j) = \sum\limits_{j=1}^s v_h(p_j) \sum\limits_{i=1}^n x_i(p_j)u_i
\end{cases}
\end{align*}
for all $\varphi_h, v_h \in V$. A generic function in $V$ can be written as the linear combination of the finite number of basis spanning $V$. This allows the solution $\hat{f}_h(p) = \vect{\psi}(p)^T \vect{\hat{f}}$ to be characterized by the linear system (\ref{eq:linear_system}) in the original paper.

\section{Simulation on the sphere}\label{app2}
\begin{figure}[H] % figuur 1
%\vspace{6pc}
\centering
\includegraphics[width=0.4\textwidth]{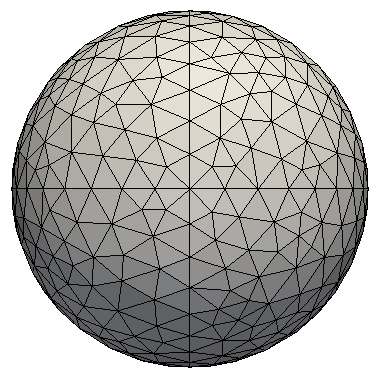}
\caption[]{The triangulated surface approximating the sphere with 488 points.}
\label{fig:mesh_s}
\end{figure}

Here we present some further simulation studies on a domain $\mathcal{M}$ that is a sphere centered on the origin and with radius $r=1$, approximated by the triangulated surface $\mathcal{M}_\mathcal{T}$ in Figure~\ref{fig:mesh_s}.

\subsection{Noisy obervations}\label{app2}
\begin{figure}[H] % figuur 1
%\vspace{6pc}
\centering
\includegraphics[width=0.9\textwidth]{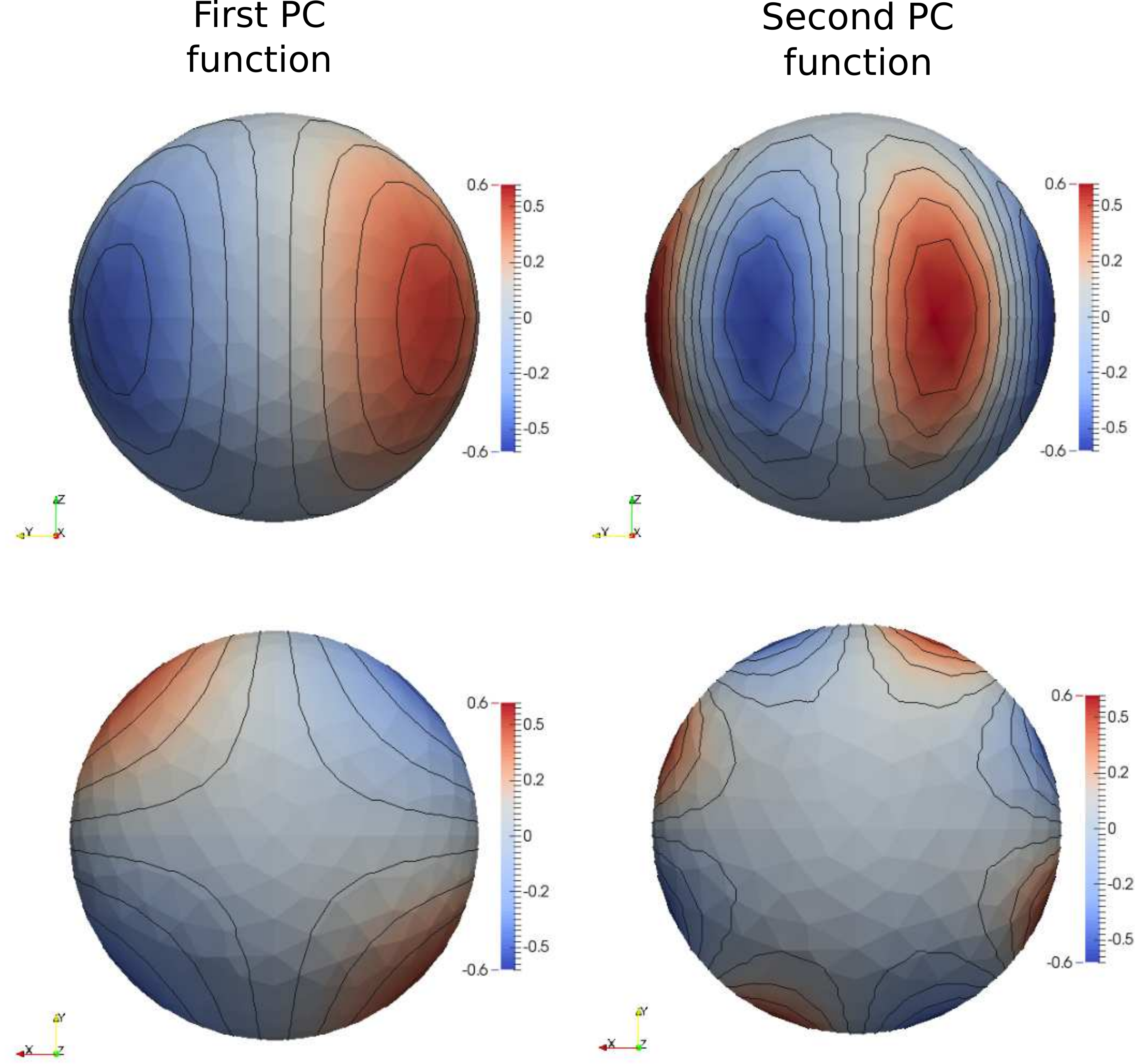}
\caption[]{From the left to the right, two views of the true first and second PC functions.}
\label{fig:original_s}
\end{figure}

We generate $n = 50$ smooth functions $x_1,\ldots,x_{50}$ on $\mathcal{M}_\mathcal{T}$ by
\begin{equation*}
x_i = u_{i1}v_1 + u_{i2}v_2, \hspace{35pt} i = 1,\ldots,n
\end{equation*}
where $v_1$ and $v_2$ represent the two PC functions with expressions
\begin{eqnarray*}
\begin{cases}
v_1(x,y,z)= \frac{1}{2} \sqrt{\frac{15}{\pi}} \frac{xy}{r^2} \\
v_2(x,y,z)= \frac{3}{4} \sqrt{\frac{35}{\pi}} \frac{xy(x^2-y^2)}{r^4}
\end{cases}
\end{eqnarray*}
and $u_{i1}$, $u_{i2}$ represent the PC scores, generated independently and distributed as $u_{i1} \sim N(0, \sigma_1^2)$, $u_{i2} \sim N(0, \sigma_2^2)$ with $\sigma_1 = 4$, $\sigma_2 = 2$. The PC functions are two components of the Spherical Harmonics basis set, so they are orthonormal on the sphere, i.e. $\int_\mathcal{M} v_i^2 = 1$ for $i \in \{1,2\}$ and $\int_\mathcal{M} v_iv_k = 0$ for $i \neq k$ with $i,k \in \{1,2\}$. The PC functions are plotted in Figure~\ref{fig:original_s}.
The functions $x_i$ are sampled at locations coinciding with the nodes of the
mesh in Figure~\ref{fig:mesh_s}. At these locations, a Gaussian white
noise with standard deviation $\sigma = 0.1$ has been added to the true function $x_i$. We are then interested in recovering the smooth PC functions $v_1$ and $v_2$ from these noisy observations.

\begin{figure}[H] % figuur 1
%\vspace{6pc}
\includegraphics[width=1\textwidth]{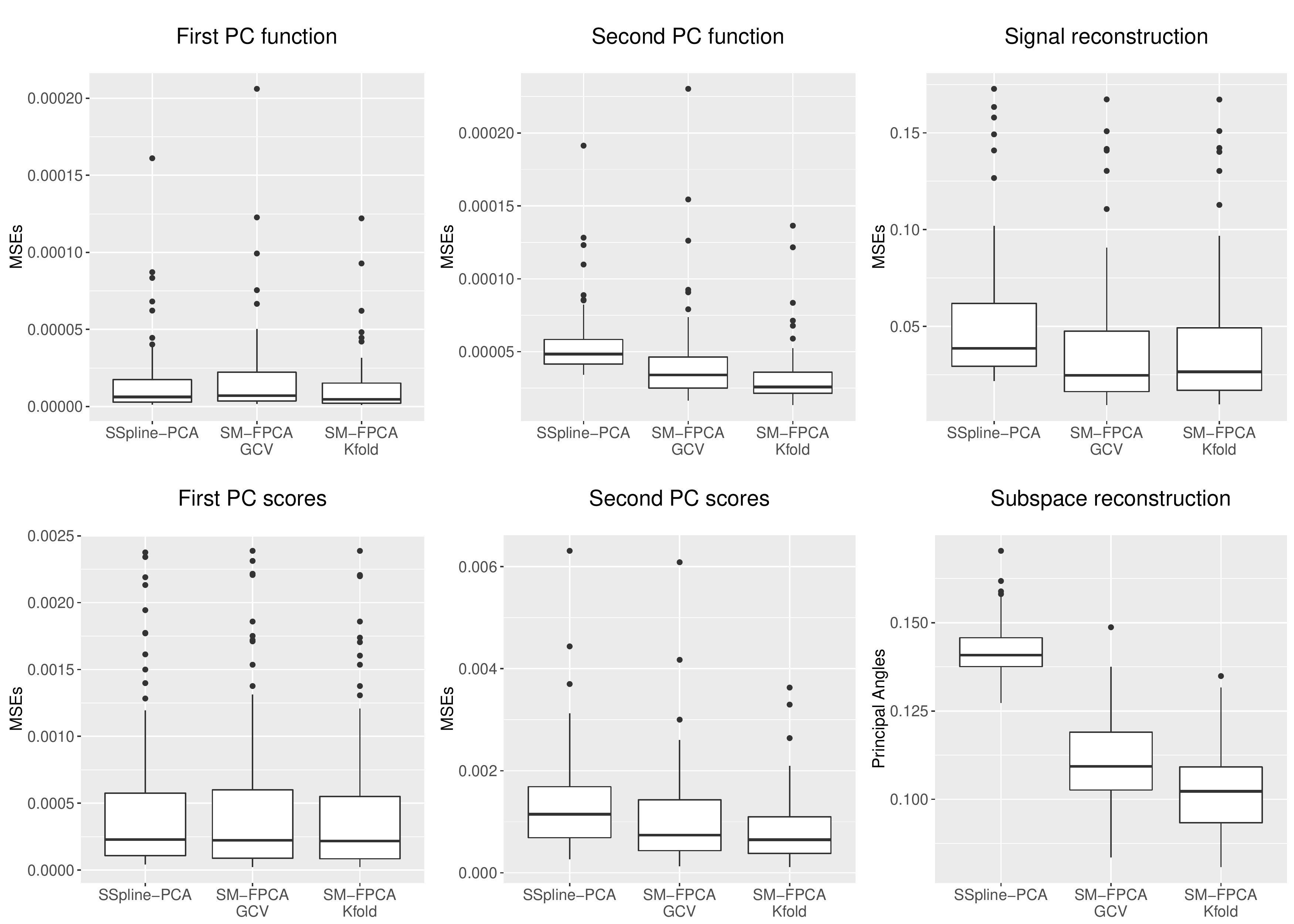}
\caption[]{Boxplots summarizing the performance of SSpline-PCA and SM-FPCA. For the SM-FPCA both GCV and $K$-fold has been applied for the selection of the smoothing parameter.}
\label{fig:boxplots_s}
\end{figure}

We apply the proposed  SM-FPCA method, choosing the optimal smoothing parameter $\lambda$, both with the $K$-fold and with GCV. We compare to the approach based on pre-smoothing followed by MV-PCA on the denoised evaluations of the functions at the locations $p_j$, $j = 1,\ldots,p$. In this case, the smoothing techniques used is Spherical Splines [\cite{wahba1981}], using the implementation in the R package \textit{mgcv}. The smoothing parameter choice is based on the GCV criterion. We will refer to this approach as SSpline-PCA. The results are summarized in Figure~\ref{fig:boxplots_s}.

The best estimates of the first two PC functions and corresponding scores are provided by the proposed SM-FPCA with selection of the smoothing parameter based on the $K$-fold approach. SSpline-PCA does a comparable job on the first principal component, but a significantly worst on the second. A possible explanation for this is the fact that SSpline-PCA tends to over-smooth the data, due to the low signal-to-noise setting of the simulations. This results in good performances for the first PC, but causes a loss of information that worsen the estimation of the second PC. Also the MSE on the signal reconstructions, as well as the measure based on the principal angle between the space spanned respectively by $\{v_i\}_{i=1,2}$ and the estimated PC functions $\{ \hat{v}_i \}_{i=1,2}$, emphasize the good performance of the introduced algorithm.

\subsection{Spatial mismatching}\label{app2}
\begin{figure}[!htb] % figuur 1
%\vspace{6pc}
\includegraphics[width=1\textwidth]{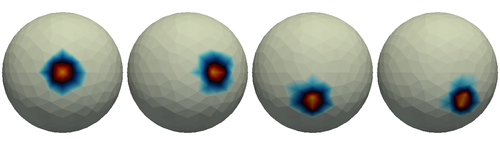}
\caption[]{A plot of the four different realizations of the misaligned PC function $v_{i1}$.}
\label{fig:mis_realizations}
\end{figure}

In this section we complement the set of simulations in the noisy setting by designing a simple simulation that shows how SM-FPCA behaves when a spatial mismatching effect is introduced. In the motivating application to neuroimaging data, spatial mismatching is introduced by the shape registration algorithm. In this simulation, we consider a spherical domain $\mathcal{M}_\mathcal{T}$ and reproduce this spatial mismatching effect, that results in misalignment of the signals on this domain, by including a subject specific shift (in spherical coordinates) of the first PC function.
In detail, we generate $n = 50$ smooth functions $x_1,\ldots,x_{50}$ on $\mathcal{M}_\mathcal{T}$ by
\begin{equation}\label{eq:gen_misaligned}
x_i = u_{i1}v_{i1}, \hspace{35pt} i = 1,\ldots,n
\end{equation}
where $u_{i1}$ represent the PC scores, generated independently and distributed as $u_{i1} \sim N(0, \sigma^2)$ with $\sigma = 4$, and the functions $v_{i1}$ represent misaligned realization of the PC function $v_1$. Specifically, we parametrize $v_1$ in spherical coordinates $(\theta, \phi)$ and set $v_{i1}(\theta, \phi) = v_1(\theta + \theta_i, \phi + \phi_i)$, with $\theta_i$ and $\phi_i$ generated independently with a discrete uniform distribution on the set $\{0, 0.4\}$. In Figure~\ref{fig:mis_realizations} we show $v_{i1}$ for the four possible realizations of shifting coefficients $(\theta_i, \phi_i)$.

The interest is to recover the structure of the only PC function $v_1$, from the misaligned realizations $\{x_i \}_{i=1,\ldots,n}$, ignoring the effects introduced by the shifts. To consider purely the misalignment's effect, we do not add noise to the sampled functions $x_i$. In fact, while the benefits of SM-FPCA in the noisy setting have already been extensively demonstrated, we aim now at considering separately the effect of a spatial mismatching on the sampled functions from the effect of the presence of noise. Pre-smoothing of the signal, as performed in SSPline-PCA, is thus unnecessary, and we compare directly MV-PCA to SM-FPCA. In fact, as already mentioned, the proposed SM-FPCA model incorporates the smoothing penalty in a more parsimonious way than the pre-smoothing approach, allowing a direct control of the smoothness of the estimated PC function. We would like to show that SM-FPCA, combined with a cross-validation approach for the choice of the smoothing parameter $\lambda$, might help removing artefacts introduced by the spatial mismatching.
\begin{figure}[!htb] % figuur 1
%\vspace{6pc}
\centering
\includegraphics[width=0.9\textwidth]{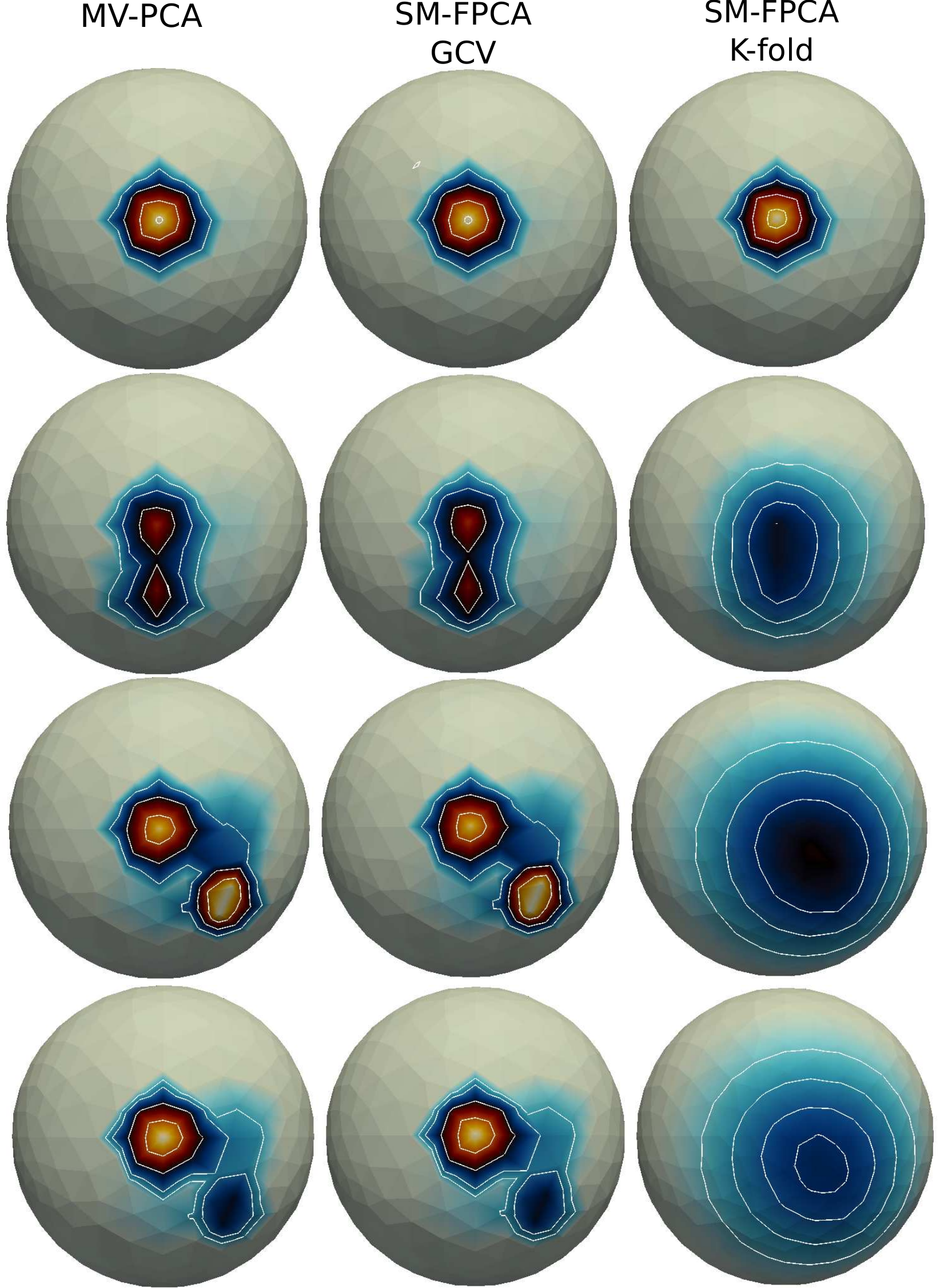}
\caption[]{From top to bottom, plot of the estimates computed on 4 different generated datasets. From left to right, plot of the estimate of the first PC function computed respectively with MV-PCA, SM-FPCA GCV and SM-FPCA K-fold.}
\label{fig:mis_estimates}
\end{figure}

In Figure~\ref{fig:mis_estimates} we show the estimates computed with MV-PCA, SM-FPCA GCV and SM-FPCA $K$-fold ($K = 5$) for four different datasets generated as in (\ref{eq:gen_misaligned}). In the top row we show a situation where the PC function estimated with MV-PCA shows a satisfactory result. In this case also SM-FPCA GCV and SM-FPCA $K$-fold show a similar behavior.  However, in the bottom three rows the estimates of the PC function computed with MV-PCA and SM-FPCA GCV show some artefacts introduced by the misalignment, while the estimate computed with SM-FPCA $K$-fold better preserves the shape of the PC function, renouncing however to spatial localization. The results obtained with SM-FPCA $K$-fold suggest to interpret the phenomena at a more macroscopical scale, due to the high local variability introduced by the spatial mismatching. 

The different behavior of SM-FPCA, when the smoothing parameter is chosen by GCV with respect to $K$-fold cross-validation, can be explained by the fact that this first approach concerns with the choice of $\lambda$ only in the regression step (\ref{eq:f_reg}), where the choice of $\lambda$ is only driven by the presence of noise on the vector $\vect{X}^T \vect{u}$. On the contrary, SM-FPCA $K$-fold is based on a direct comparison of the PC function estimated on the training and validation sets, obtained partitioning the dataset.

\section*{Acknowledgements}

JADA would like to acknowledge support from the Engineering and Physical Sciences Research Council (Grants EP/K021672/2, EP/N014588/1). The authors would very much like to thank the Editor, Associate Editor and Referees for their really encouraging comments throughout the process.

\bibliographystyle{abbrvnat}

\bibliography{Bibliography} % The references (bibliography) information are stored in the file named "Bibliography.bib"

\end{document}